\documentclass[apj]{emulateapj}

\usepackage{amsmath}

\let\ol\relax
\newcommand{\chisq}{\ensuremath{\chi^2}}
\newcommand{\etal}{et~al.}
\newcommand{\om}{\ensuremath{\Omega_{m}}}
\newcommand{\ol}{\ensuremath{\Omega_{\Lambda}}}

\newcommand{\zhel}{\ensuremath{z_{\mathrm{hel}}}}
\newcommand{\gmeg}{\ensuremath{g_M}}
\newcommand{\rmeg}{\ensuremath{r_M}}
\newcommand{\imeg}{\ensuremath{i_M}}
\newcommand{\zmeg}{\ensuremath{z_M}}
\newcommand{\scriptc}{\ensuremath{\mathcal{C}}}
\newcommand{\snlsfilts}{\ensuremath{\gmeg \rmeg \imeg \zmeg}}
\newcommand{\sint}{\ensuremath{\sigma_{\mathrm{int}}}}

\begin{document}

\title{SiFTO: AN EMPIRICAL METHOD FOR FITTING SN~Ia LIGHT CURVES$^1$}
\shorttitle{SiFTO}
\shortauthors{Conley \etal}

\author{
 A.~Conley\altaffilmark{2},
 M.~Sullivan\altaffilmark{2,3},
 E.~Y.~Hsiao\altaffilmark{4},
 J.~Guy\altaffilmark{5},
 P.~Astier\altaffilmark{5},
 D.~Balam\altaffilmark{4},
 C.~Balland\altaffilmark{5},
 S.~Basa\altaffilmark{6},
 R.~G.~Carlberg\altaffilmark{2},
 D.~Fouchez\altaffilmark{7},
 D.~Hardin\altaffilmark{5},
 D.~A.~Howell\altaffilmark{2},
 I.~M.~Hook\altaffilmark{3},
 R.~Pain\altaffilmark{5},
 K.~Perrett\altaffilmark{2},
 C.~J.~Pritchet\altaffilmark{4},
 N.~Regnault\altaffilmark{5}
}

\altaffiltext{1}{
 Based on observations obtained with MegaPrime/MegaCam, a joint project
 of CFHT and CEA/DAPNIA, at the Canada-France-Hawaii Telescope (CFHT)
 which is operated by the National Research Council (NRC) of Canada,
 the Institut National des Sciences de l'Univers of the Centre National
 de la Recherche Scientifique (CNRS) of France, and the University of
 Hawaii. This work is based in part on data products produced at the
 Canadian Astronomy Data Centre as part of the Canada-France-Hawaii
 Telescope Legacy Survey, a collaborative project of NRC and CNRS.
}
\altaffiltext{2}{Department of Astronomy and Astrophysics, University
  of Toronto, 50 St. George Street, Toronto, ON M5S 3H4, Canada}
\altaffiltext{3}{University of Oxford Astrophysics, Denys Wilkinson
  Building, Keble Road, Oxford OX1 3RH, UK}
\altaffiltext{4}{Department of Physics and Astronomy, University of
  Victoria, PO Box 3055, Victoria, BC V8W 3P6, Canada}
\altaffiltext{5}{LPNHE, CNRS-IN2P3 and University of Paris
  VI \& VII, 75005 Paris, France}
\altaffiltext{6}{LAM CNRS,
  BP8, Traverse du Siphon, 13376 Marseille Cedex 12, France}
\altaffiltext{7}{CPPM, CNRS-IN2P3 and University Aix Marseille II,
  Case 907, 13288 Marseille Cedex 9, France}

\email{ conley@astro.utoronto.ca }

\begin{abstract}
We present SiFTO, a new empirical method for modeling Type~Ia
supernovae (SNe~Ia) light curves by manipulating a spectral template.
We make use of high-redshift SN data when training the model,
allowing us to extend it bluer than rest frame $U$.  This increases
the utility of our high-redshift SN observations by allowing us to use
more of the available data.  We find that when the shape of the light
curve is described using a stretch prescription, applying the same
stretch at all wavelengths is not an adequate description.  SiFTO
therefore uses a generalization of stretch which applies different
stretch factors as a function of both the wavelength of the observed
filter and the stretch in the rest-frame $B$ band.  We compare SiFTO
to other published light-curve models by applying them to the same set
of SN photometry, and demonstrate that SiFTO and SALT2 perform better
than the alternatives when judged by the scatter around the best fit
luminosity distance relationship.  We further demonstrate that when
SiFTO and SALT2 are trained on the same data set the cosmological
results agree.
\end{abstract}


\section{INTRODUCTION}
\label{sec:introduction}
\nobreak 
The importance of modeling Type~Ia supernova (SN~Ia) light curves is
clear from the large number of methods that have been developed
for this purpose.  An incomplete sample includes MLCS/MLCS2k2
\citep{1996ApJ...473...88R, 2007ApJ...659..122J}, stretch
\citep{1997ApJ...483..565P, 2001ApJ...558..359G, 2003ApJ...598..102K},
super stretch \citep{2006ApJ...641...50W},  $\Delta m_{15}$ 
\citep{1996AJ....112.2438H, 2006ApJ...647..501P}, BATM
\citep{2003ApJ...594....1T}, CMAGIC \citep{2003ApJ...590..944W}, SALT
\citep{2005A&A...443..781G}, and SALT2 \citep[hereafter
G07]{2007A&A...466...11G}.  For some SNe the different techniques 
produce quite different results, but when applied to a moderately sized
sample the overall results appear to be fairly similar
\citep{2007ApJ...666..694W}.  These packages are most frequently used
to derive relative distances which can then be used to constrain the
cosmological parameters.  Depending on the data set, current methods
give relative distances to a precision of $\sim 7-10\%$.  The potential
reward for any small improvement in how well distances can be
extracted from SN data is large.

This paper presents the details of SiFTO, a light-curve analysis tool
developed for use with data from the Supernova Legacy Survey
\citep[SNLS;][hereafter A06]{2006A&A...447...31A}.  SNLS follows a
two-pronged analysis approach: most of the steps are carried out
twice using independent methods, and the results are compared.  This
includes spectroscopic typing, photometry, calibration, light-curve
fitting, and the extraction of the cosmological parameters.  The
details of the complementary light-curve fitting method (SALT/SALT2)
are described in \citet{2005A&A...443..781G} and G07.  By design,
SiFTO takes a simpler algorithmic approach than SALT2, which should
make the results somewhat more robust to the details of the training
sample.  SALT2, however, should have more growth potential as future
data sets become available.

The features of SiFTO are:
\begin{itemize}
  \item The shape of the light-curve is described by a single stretch
    parameter, the stretch in the rest-frame $B$ band, $s_B$.
    However, the model is generalized so that different observed
    filters stretch by different amounts as a function of 
    wavelength and $s_B$ (\S~\ref{subsec:stretch}).
  \item SN~Ia colors are handled by allowing the normalization of the fit
    in each observed filter to vary independently.  The resulting colors are
    then combined to form a single color parameter when the results of
    the light-curve fit are transformed into a distance
    (\S~\ref{subsec:colormodel}).
  \item For the A06 data sample, the RMS of the SiFTO-derived distances
    around the best fit cosmological parameters is 0.16 mag.  This is
    identical to SALT2 on the same data set, and is a clear improvement
    over SALT and MLCS2k2 (\S~\ref{subsec:hizcomparison}).  
  \item When we train SiFTO and SALT2 on the same data set using the
    same non-SN related settings (filter functions, etc.) the cosmological
    results are very similar (\S~\ref{sec:retrain_compare}).
\end{itemize}
The outputs of SiFTO are a set of light-curve parameters (the time of
maximum and the $s_B$) and the spectral energy
distribution (SED) of the SN as a function of time.  From this
information one can construct a distance estimate by finding the
magnitude in the rest frame $B$ band ($m_B$) on a fiducial epoch
and some set of rest-frame colors.  The peak magnitude, stretch, and
color can be combined to form a relative distance estimate.

There are many similarities between SiFTO and SALT/SALT2.  Unlike the
other packages mentioned above, SiFTO, SALT, and SALT2 work in flux space by
manipulating a model of the SED directly.  By working with
the SEDs, the fits are naturally performed in the observer frame
without the need for a distinct $K$-correction step.  Like SALT2,
SiFTO incorporates high-redshift SNLS data when deriving the model SED, which
allows us to extend our model further towards the blue than is
possible with only low-z data.  This is particularly important when
working with surveys that sample a large range of redshifts, since in
this case some of the observed filters may be bluer than the
rest-frame $U$ band at the highest redshifts.  This is true even of
many surveys which obtain supplemental near-IR photometry.

SNLS observes SNe in the range $0.2 < z < 1$ in four filters (\snlsfilts
), which are similar to the USNO system described in
\citet{2002AJ....123.2121S}.  At the lowest-redshifts, these filters
essentially sample $BVR$, and at highest $\imeg \mapsto U$ and $\zmeg
\mapsto B$.  It is critical that our model handles both situations in
a consistent fashion.  For $z > 0.55$, \gmeg\ is bluer in the rest
frame than $U$, and the same holds for \rmeg\ at $z > 0.9$.  Since
the SN is often still detected in these filters, if our model extends
only to $U$ (3600 \AA ) we will have to ignore some of our signal.
Incorporating \rmeg\ for the highest redshift SNe is particularly
helpful because the \zmeg\ data are both noisier (due to fringing, sky
brightness, and a drop in the CCD quantum efficiency) and harder to
calibrate than \gmeg , \rmeg , and \imeg , and it is necessary that we have
good quality observations of each SN in at least two filters to
measure the color.

There are many algorithmic differences between the fitters.  SiFTO 
uses a spectral template as an input to the training process, although
this template was informed by the results of initial SiFTO fits, while
SALT2 includes spectra more directly in the model derivation.  The
method of adjusting the SED models are also very different, with SALT2
in principle allowing for more detailed modeling of individual
features.  However, given the limitations of current training samples,
the most important difference between SALT/SALT2 and SiFTO lies in how
SN colors are handled.  In SiFTO, the SED is adjusted to match the
observed frame colors in each filter, rather than by imposing a single
color model across all filters simultaneously as in SALT/SALT2.  The
justification is that SNe~Ia show evidence for some additional
variability in their colors, particularly in the near-UV, so forcing
the fit to match a single color law at all wavelengths may give a poor
result.  Note that SiFTO, SALT, and SALT2 do account for a
relationship between stretch and color when the results are converted
into a distance.  In this paper, this relationship is taken to be
linear.

SiFTO is designed for use with modern, well measured SN data sets, and
is not suited for the analysis of poorly sampled light curves.
Specifically, if the number of observations in a filter in the
range -20 to 40 rest frame days relative to $B$ maximum is less than
about 3, then the fit in that filter may not be reliable.  This
proviso only applies to historical SN~Ia samples -- even at the
highest redshifts, current observations are of high enough quality
that this is not a problem.  Our approach is not well suited to the
rest-frame near-IR ($I$ or redder), where a stretch-like prescription
does not work well.

Our base SED model is an updated version of \citet[hereafter
H07]{2007ApJ...663.1187H}, adjusted to match SN~Ia observations in a
number of synthetic filters (\S~\ref{subsec:sedmodel}).  Our template
includes an error snake representing the uncertainty in the template
and intrinsic variability of SNe~Ia
(\S~\ref{subsec:errorsnake}). SiFTO is designed so that the effects of
most systematics can be modeled by adjusting the color relation, and
do not require fully retraining the entire template
(\S~\ref{subsec:systematics}).  The data used to train the SiFTO
model is described in \S~\ref{sec:data}.  The derived light-curve
parameters are the most reliable when there are data within $\sim 7$
rest frame days of the epoch of peak $B$ luminosity
(\S~\ref{sec:daygap}).

\section{APPROACH}
\label{sec:approach}
\nobreak

\subsection{Light-curve fitters and distance estimators}
\label{subsec:lcvsdistance}
\nobreak 
The distinction between light-curve fitters and distance estimators is
often not made explicit in the literature. The majority of the
published packages are light-curve fitters, with the exception of
MLCS/MLCS2k2 and BATM.  A light-curve fitter attempts to find the best
fit to a given set of SN~Ia photometry.  The parameters of this fit
can then be converted into a distance estimate if desired, but technically
this step is not part of the fit.  A distance estimator attempts to
find the distance directly rather than trying to obtain the best fit
to the data. In both cases one is only trying to find relative
distances; nothing described in this paper depends on knowing the absolute
distance to any SN, the Hubble constant, or the absolute magnitude of
any SN in any bandpass.

The advantage of a distance estimator is that the output is directly
what is desired for {\it most} applications of SN~Ia data.  Therefore,
the products of such an analysis are simpler to use, and in principle
such an approach may do a better job extracting the information
directly relevant to this purpose. Their primary disadvantage is that,
by their nature, they must use distance information in
their training, usually in the form of residuals from the best-fit
Hubble relation.  This makes it difficult to include both very
low-redshift SNe (which are not in the smooth Hubble flow), and
high-redshift data (where the residual depends on the cosmological
parameters).  To use high-z data properly, it would be
necessary to re-train the model from scratch for every value of the
cosmological parameters one evaluates, which would be extremely
computationally expensive.  Therefore, in practice, light-curve
fitters have access to a more data for training purposes.  Neither
approach is obviously superior, and we will not attempt to give a
comprehensive comparison of their merits here.  Because we find the
advantages of incorporating data from a range of redshifts in our
training compelling, SiFTO is a light-curve fitter.

\subsection{SED modeling}
\label{subsec:modeltype}
\nobreak 
Like SALT/SALT2, SiFTO works by manipulating a SED model.
Most fitters do not take this approach.  Instead, they are defined in
some set of rest-frame filters and use $K$-corrections to
transfer the model to the observed frame.  That is, they work
in photometry space rather than SED space.  This is simpler,
but the fact that the $K$-corrections must be calculated and applied
in a distinct step can lead to complications in ensuring that the
two steps are fully consistent with each other.  Working
directly with the SEDs obviates this difficulty; formally, SiFTO,
SALT, and SALT2 do not calculate $K$-corrections, although this
is perhaps more of a pedagogical distinction than a practical one.  This
choice requires that we work with fluxes rather than magnitudes, which
is desirable anyways because using magnitudes will bias fits to
low signal-to-noise data.

We start with a base model for some fiducial SN~Ia and then use
additional parameters to adjust the SED to best match a particular SN.
Two parameters are fairly obvious (some overall flux normalization and
overall time offset), but beyond this the parameterization is somewhat
arbitrary.  Theoretical modeling of SNe~Ia is sufficiently difficult
that it offers little guidance here, so our choice of parameters is
empirical.  The existence of a relation between between the shape of
the light curve and the peak brightness of the SN is very well
established, as is the relation between the color and luminosity.
These are in the sense that more slowly evolving SNe are observed to
be more luminous (the wider-brighter relationship), and redder SNe are
fainter (bluer-brighter).  The former is clearly related to SN
physics, while the latter is very likely some combination of SN
properties and external effects such as dust
\citep{1992ARA&A..30..359B,2007ApJ...664L..13C}.

\subsection{Light-curve shape}
\label{subsec:stretch}
\nobreak
The SiFTO model for light-curve shape is based on the stretch
parameterization \citep{1997ApJ...483..565P,2001ApJ...558..359G}.
Here, the timescale over which a SN~Ia evolves is modified by multiplying
by a factor $s$.  SNe with higher $s$ evolve more slowly, with broader
light curves.  Changing the stretch simply widens or narrows the
light curve in a linear fashion.  

In most previous applications of stretch either the same value was
applied at all wavelengths or the stretch values were fit
independently in each filter.  The quality and quantity of SNLS data
shows that the first is no longer sufficient, and we can improve on
the second by relating the stretches of different filters.  SiFTO uses
an empirical relation between the stretch in $B$ ($s_B$) and the
effective rest-frame wavelength of the observed filters. This is
constrained to work like a simple stretch factor in rest-frame $B$.
With this adjustment, the stretch model is a very good description of
our observations, as will be demonstrated later.  However, this
approach does not work well in the rest-frame $I$ and redder, which
may be an issue for some data sets, although not SNLS. The SALT model
is based on the stretch prescription in $B$, but not in other filters,
while SALT2 allows for more general behavior.

\subsection{SN colors}
\label{subsec:colormodel}
\nobreak 
Rather than imposing a relation between different observed filters
during the light-curve fit, in SiFTO this relation is used when the
fit parameters are converted into a distance.  The normalization of
the fit to each observed filter is allowed to float independently
during the fit.  The SED is then adjusted using splines to match these
observed colors.  We find that single parameter functions (such as
dust laws or the SALT/SALT2 color relation) are generally unable to
reproduce the full variety of observed colors across more than two
observed filters, which is likely indicative of some intrinsic
variability in SNe~Ia.

Therefore, the parameters in each fit are the time of maximum
luminosity in the rest-frame $B$-band $t_0$, the stretch $s_B$, and $N_f$
normalization parameters, one for each observed filter whose rest
frame wavelength lies within the range our model is defined.  $N_f = 4$ for
most SNLS data, except at the redshift extremes.  The results can be
used to produce a corrected distance which makes a better standard candle:
\begin{equation}
 m_{corr} = m_B + \alpha \left( s_B - 1 \right) - \beta \scriptc
 \label{eqn:mbcorr} .
\end{equation}
$\alpha$ and $\beta$ must be determined empirically along with
the cosmological parameters, $m_B$ is the peak rest-frame magnitude in
the $B$ band, and \scriptc\ is some sort of color parameter.  In
principle we could have several color parameters, one for each
observed color, but it would be difficult to ensure that the color was
being applied consistently across the entire sample.  We therefore combine the
different observed colors into a single parameter, using the SN data
themselves to derive the relationship between different rest-frame colors.

One of the consequences of this relation is that any linear variation
of the intrinsic color of SNe~Ia with stretch is absorbed by the
$\alpha$ term.  Therefore, our method {\it does} model some aspects of
intrinsic color and how it relates to light-curve shape.  The
differences between this approach and that currently used by MLCS2k2
are threefold: First, MLCS2k2 includes a quadratic term in the
light-curve shape.  This has a small effect except at the extremes of
the SN population.  Second, MLCS2k2 makes use of several priors on the
parameters, of which the color prior is the most important.  Third, it
assumes that any residual color that remains after the correction for
light-curve shape is due to Milky Way-like dust, and therefore $\beta$
is (usually) set to 4.1 for the $B-V$ color. In effect, these
assumptions mean that SN colors can only derive from three sources:
the shape of the light curve, dust, and possibly some scatter in the
intrinsic colors which has no effect on the luminosity.  MLCS2k2 could
be modified to relax this set of assumptions.  Therefore, the current
distinction between intrinsic and extrinsic colors in MCLS2k2 versus
SiFTO and SALT/SALT2 is more practical than philosophical, and is
mostly related to the exact nature of the color and light-curve shape
priors used by MLCS2k2.  Whether or not these priors are beneficial
depends on whether the smaller statistical errors obtained by using
them outweigh the enhanced sensitivity to evolution or other
systematic errors in the prior distributions (e.g., the distribution
of extinction and how it interacts with survey selection effects) that
they introduce.

\citet{2007ApJ...664L..13C} demonstrates that either dust along the
line of sight to local SNe is either quite different than we expect,
or that SN intrinsic colors are not predicted completely by a linear
or quadratic light-curve shape model by showing that $\beta$ is
significantly less than 4.1.  We have further investigated this
question by introducing a quadratic stretch term into our corrected
magnitudes, and find that it has virtually no effect on $\beta$.  In
other words, adding an additional quadratic term like the one used in
MLCS2k2 does little to address the issue of intrinsic versus extrinsic
color.

We build a model that can combine the different measured colors for
individual SNe into a single color parameter by taking SNe with
simultaneous rest-frame $UBV$ observations and using this data to
predict $B-V$ as a function of $U-B$ and $s_B$.  Here, as elsewhere in
this paper, colors are always measured at the epoch of $B$ maximum.
In effect, this is loosely similar to the SALT/SALT2 approach, except
that the color model has been separated from the light-curve fit and
the SED model building.  In A06 a similar relation, implicitly
included in the SALT training process, was determined purely by low-z
SN data.  $U$ data is notoriously difficult to calibrate
\citep[e.g.,][]{1990PASP..102.1181B}, and relatively little is available,
although the situation has improved considerably recently
\citep{2007ApJ...659..122J}.  By using SNLS data we can overcome this
problem -- the rest-frame $U$ generally maps to either observer frame
$\gmeg$ or $\rmeg$, both of which are well calibrated.  In addition,
this data is available for virtually all of our SNe.  As a result, the
color relation is considerably better constrained than was the case in
A06.

Our color relation takes the form
\begin{equation}
 \left(B - V\right)_{U-B,s} = a \left( U - B + 0.5 \right)
  + b \left( s_B - 1 \right) + c .
 \label{eq:colorrelation}
\end{equation}
We then combine this with the actual measured $B-V$ (if available)
using a weighted average to form our color parameter \scriptc , which
therefore represents the $B-V$ color of the SN at peak, usually after
correction for Milky-Way extinction.  This differs by an arbitrary
constant offset from the definition used in SALT/SALT2.

The data suggests that there is some additional intrinsic dispersion
in how well $B-V$ can be predicted by $U-B$ beyond that attributable
to measurement errors -- in other words, the different colors do carry
some independent information.  This should not be entirely surprising,
as there has long been a theoretical expectation that SNe~Ia are not
as homogeneous in the near-UV as at longer wavelengths
\citep{2000ApJ...530..966L}. This is supported by observations
\citep{2007ellis}.  We account for this by degrading the assigned
uncertainty in the predicted $B-V$ by an additional factor
$\sigma_{\mathrm{clr}}$ added in quadrature with the measurement
error.  This plays a similar role to the $K$-correction uncertainties
discussed in G07, although it is expressed in a different fashion.

Because of this additional scatter it is important that the statistic
used to measure $a, b, c$ allows for dispersion beyond that
represented by the measurement errors.  We use the multiple regression
method of \citet{2007ApJ...665.1489K}, which also provides estimates
of $\sigma_{\mathrm{clr}}$, and couple these fits with iterative
$2.5 \sigma$ outlier rejection when deriving the relation.

A further refinement is to extend this model bluer than the rest-frame
$U$-band.  We define an artificial filter defined as the $U$-band
filter of \citet{1990PASP..102.1181B} blueshifted
by $z=-0.2$, which we refer to as $U_{02}$, and which has an
effective wavelength of $\sim 3000$ \AA.  We then use a similar
relation to the above to predict $B-V$ given $U_{02}-B$. Not
surprisingly, this relationship is not as well determined since it is
measured with higher-redshift, and hence noisier, observations.
Nonetheless, the additional information provided by $U_{02}$ is useful
with the highest-redshift SNLS observations (\S \ref{sec:results}).
We denote the coefficients of this relation as $a_{02}\, b_{02}\,$ and 
$c_{02}$, with a corresponding $\sigma_{02\mathrm{clr}}$.  The same
technique can also be applied to $V-R$.  Since this is not useful for
the majority of SNLS data (which do not have rest frame $R$ observations), 
we do not discuss these coefficients here.

This gives us (potentially) three estimates of $B-V$ -- that from
$U-B$, $U_{02}-B$, and finally the actual $B-V$.  We form a
weighted mean of whatever subset are available to form \scriptc ,
taking care to propagate the covariances.  The additional scatter
represented by $\sigma_{\mathrm{clr}}$ and similar terms
usually has a significant effect on the final weights.  Large
extrapolations are undesirable in this process because they increase
the importance of the boundary conditions used when we adjust the SED.
To avoid this, we require that each of our synthetic rest frame
magnitudes ($U_{02}UBV$) has an observed frame filter within 650 \AA\
(in the rest frame), and do not allow the same observed filter to be
matched to multiple rest-frame filters.  With the requirement in place,
our fits are very insensitive to the form of the boundary conditions.

The model blue-ward of $U$ ($\sim 3600$ \AA ) is completely determined
by SNLS data.  This means that the question of how well this part of
the model is calibrated observationally is determined by our \gmeg\
and \rmeg\ calibrations, which are secure.  The translation from
magnitudes to flux is not as precise in this wavelength regime (i.e.,
we do not know the flux of Vega in $U_{02}$ as well as in the standard
filters), and so in an absolute sense our model SED has some
additional calibration uncertainties here.  However, because we do not
have any observer frame data in this wavelength range, these are
completely irrelevant to our current application.  That is, the SED of
Vega could be replaced with any function below 3000 \AA\ and it
would have no effect on our derived distances.  Of more interest is
the concern that SNe~Ia may not be good standard candles in the
near-UV, and that there may be external factors (such as dust) which
vary more with environment at 3000 \AA\ than at 3600 \AA .

\section{DETAILS OF THE MODEL}
\label{sec:details}
\nobreak 
Given our SED model $\phi$ (described in \S~\ref{subsec:sedmodel}),
parameterized by the observed epoch $t$, the stretch in the $B$ band
$s_B$, the epoch of peak luminosity in the $B$ band $t_0$, the $N_f$
phase-independent multiplicative factors $n_i$, we define the
effective broadband integrated flux $F_j$ in a filter $j$ of a SN
observed at a heliocentric redshift \zhel\ by
\begin{equation}
\label{eq:template_integral}
F_j =\frac{\int_0^{\infty}
  \phi \left( \frac{\lambda}{1+\zhel},t,t_0,n_i,s_B \right)
   T_j\left( \lambda \right) \lambda \mathrm{d}\lambda }
   {\left( 1+ \zhel \right) \int_0^{\infty}
   T_j\left( \lambda \right) \lambda \mathrm{d} \lambda },
\end{equation}
where $T_j(\lambda)$ is the dimensionless total (atmosphere + optics +
filter + CCD) system transmission function. Note that this is not the
energy flux absorbed by the detector, differing both in normalization
and by including an extra factor of $\lambda$ inside the integral.
The utility of this quantity stems from its relation to astronomical
magnitudes: $m = -2.5 \log_{10} F_j + \mathrm{const}$.  The
factor of $\lambda$ is a result of the fact that
astronomical magnitudes are defined in terms of
counts, not energy.  The calibration process by which observations are
placed on a standard photometric system involves a similar term over
some reference spectrum whose magnitudes are known; this is
conventionally taken to be the spectrum of Vega.  We require the $n_i$
to be positive.  The $T_j$ are never known perfectly, and this should
be included in the systematics error budget of any cosmological
analysis, including the consequences for SiFTO or any other
light-curve fitter.  An exploration of how well these are known for
different supernova samples is beyond the scope of this paper.  We
expect to update SiFTO as our understanding of various calibrations
improves and new SN samples become available.

The light-curve fit then minimizes
\begin{equation}
\label{eq:lcfit}
\chisq = \sum^{N_f}_{j=1}\sum^{N_{j}}_{i=1}\frac{\left(
 F_j(t, t_0, n_i, s_B, \zhel )-f_{ij}\right)^2}{\sigma^2_{ij}+
 \sigma_F^2 \left(t, t_0, n_i, s_B, \zhel,\lambda^{j}_{\mathrm{eff}}\right)}
\end{equation}
where $N_{j}$ is the number of datapoints in the j$^{\mathrm{th}}$
filter, $\sigma_F$ is the template error, $\lambda^j_{\mathrm{eff}}$
is the effective wavelength of filter $j$ (see below), and $f_{ij}$
are the observed data points with errors $\sigma_{ij}$.

The parameterization of $\phi$ is the most critical element.  We start
with a model for the SED of a fiducial $s_B=1$ SN~Ia, $\phi_0\left(
\lambda, t \right)$, where we define $s_B$ so that the mean value
averaged over our training sample is unity, and $\phi_0$ is a time series
of spectral templates covering the range 2000--9000 \AA\ with a unique
SED at every epoch from $-19$ to $+70$ days.  We then form the effective
epoch $\tau = \left( t - t_0 \right) / \left(s_B \left(1+ \zhel
\right) \right)$, and then use this quantity to select the appropriate
SED from $\phi_0$, linearly interpolating between the two
nearest tabulated epochs.

We next apply the stretch model as described below, and
adjust the SED to match the observed colors so that the final
result is a single SED which can be used to predict the magnitudes in
any filter at any desired redshift.  Usually this is used to compute
rest-frame magnitudes in some standard set of filters such as
$U_{02}UBV$.

\begin{figure*}
\plotone{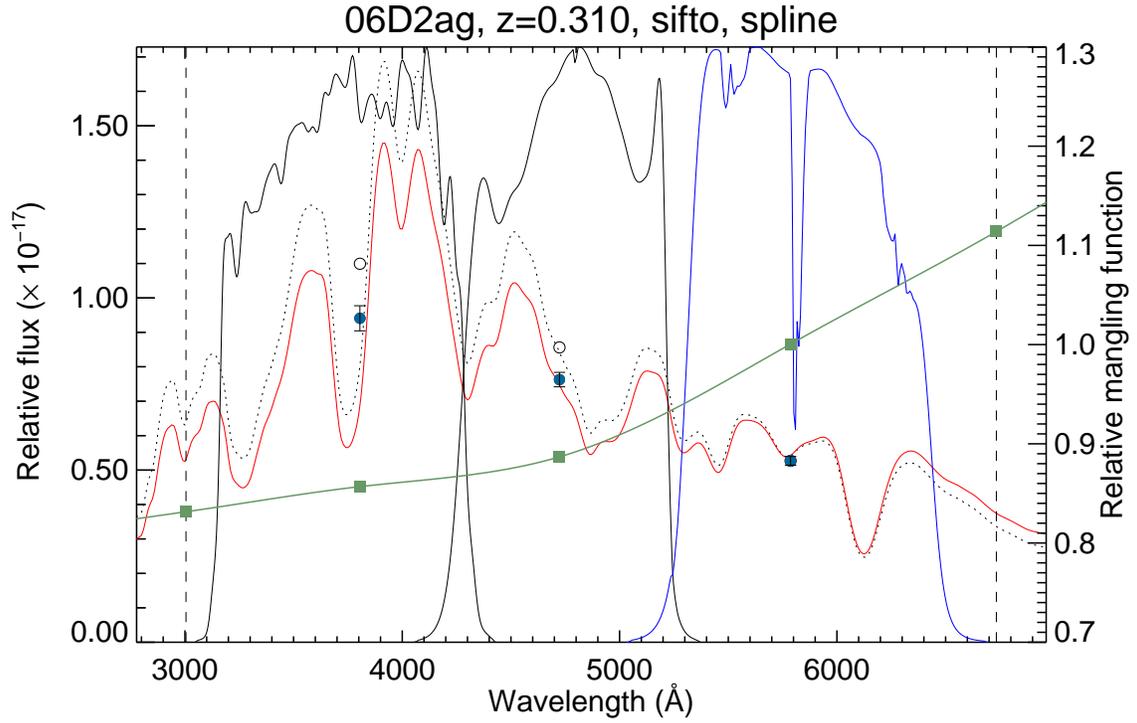}
\caption[Example of SED adjustment]{Example of adjusting the SED to
 have the desired observer frame colors using the relatively low-z SNLS SN 
 06D2ag ($z=0.310$).  Here \zmeg\ is not used because it is redder than
 the range over which the model is trained.  The dotted line is the fiducial
 template $\phi_0$, the solid red line is the adjusted SED, the smooth
 line is the warping spline, the open circles are the original integrated
 flux through each of the observed filters (also shown), and the filled
 circles are the final, target fluxes.  The rest-frame magnitudes are evaluated
 by performing synthetic photometry on the SED.  The wavelengths are
 given in the rest frame. \label{fig:mangle_example} }
\end{figure*}

The method used to adjust the SED is similar to that described in H07:
we multiply the SED by an interpolating cubic spline under tension
($\mathcal S$), with the spline knots placed at the effective
wavelengths of the observed filters.  In
order to ensure that the SED remains positive, we work in log space.
An example is shown in figure~\ref{fig:mangle_example}.  We
experimented with a variety of functional forms to smoothly scale the
spectrum in place of splines, and found similar results using linear
or higher-order polynomial interpolation.  

\subsection{The stretch model}
\label{subsec:stretchmodel}
\nobreak
In the SiFTO model, $s_j$, the stretch in filter $j$, is not
necessarily the same as $s_B$.  Our model for this correction is $s_j
= h \left( s_B, \lambda^B_{\mathrm{eff}} - \lambda^j_{\mathrm{eff}}
\right)$, where $\lambda^B_{\mathrm{eff}}$ and
$\lambda^j_{\mathrm{eff}}$ are the effective rest frame wavelengths of
$B$ and the $j^{\mathrm{th}}$ filter for our fiducial, $s_B=1$ SED at
$t=t_0$:
\begin{displaymath}
  \lambda^B_{\mathrm{eff}} = \frac{ \int \phi_0 \left( \lambda, t = t_0 \right)
       T_B \left( \lambda \right) \lambda^2 \mathrm{d} \lambda }
                   { \int T_B \left( \lambda \right) \lambda \mathrm{d} 
                     \lambda } ,
\end{displaymath}
and similarly for $\lambda^j_{\mathrm{eff}}$, although for the latter
the redshift of the SN needs to be taken into account.  
Given the stretch in the observed filter, we adjust the corresponding flux
normalization by multiplying it by the ratio:
\begin{displaymath}
  \mathcal{R}_{j} \left( t, t_0, s_B, s_j, \zhel \right) = 
    \frac{ \int \phi_0 \left( \lambda, \frac{t - t_0}
     {s_j \left(1 + \zhel \right)} \right) T_j \left( \lambda 
    \right) \lambda \mathrm{d} \lambda }
       { \int \phi_0 \left( \lambda, \frac{t - t_0}{s_B \left(1 + 
           \zhel \right)} \right) T_j \left( \lambda 
         \right) \lambda \mathrm{d} \lambda } .
\end{displaymath}
This gives observer-frame flux in each observed filter.  There
is still only one light-curve shape parameter, $s_B$.

\begin{figure*}
\plotone{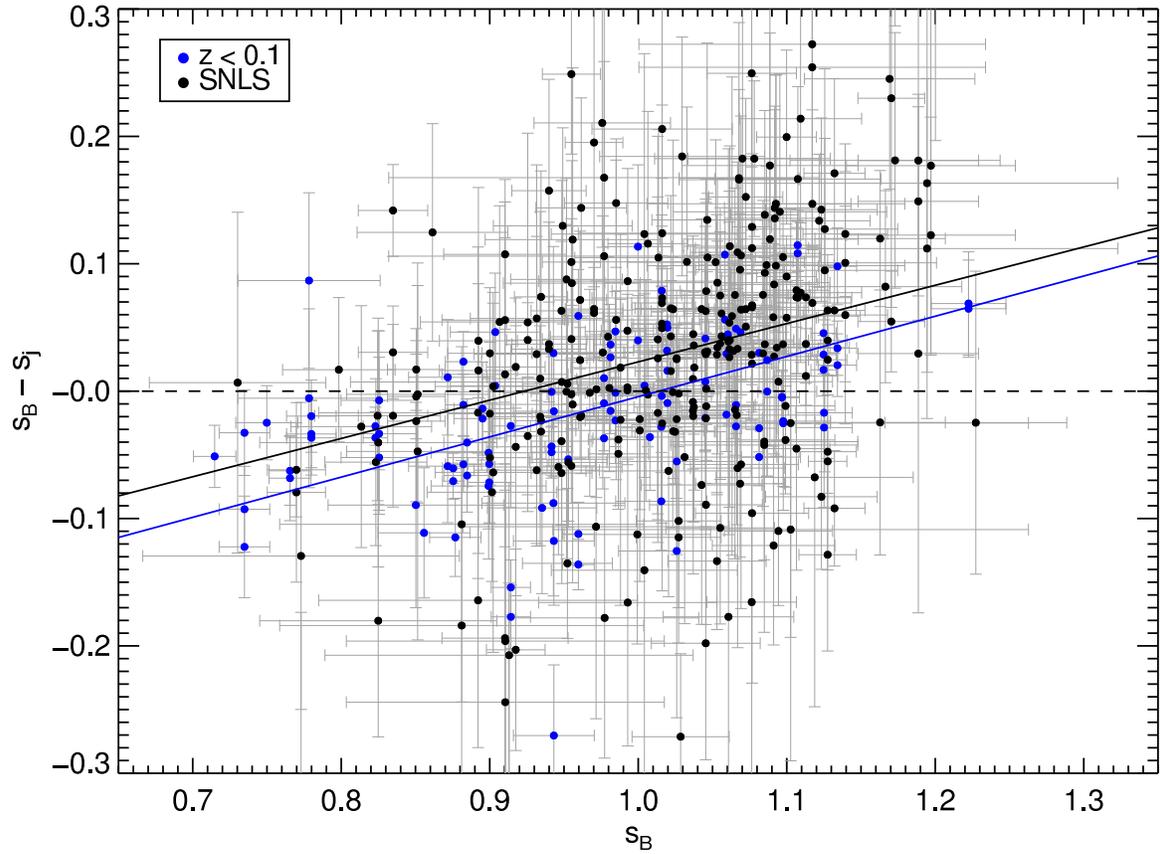}
\caption[Difference between $s_B$ and $s_j$ as a function of
 $s_B$] {Difference between $s_B$ and $s_j$ as a function of $s_B$
  for all the SNe in the training sample.  SNLS SNe~Ia are shown as
  black circles, low-z SNe~Ia as blue circles.  The best fitting linear
  model (including errors in both $s_B$ and $s_j$) for both data sets
  is over-plotted.\label{fig:stretchratio_sB} }
\end{figure*}

The stretch function $h$ is represented by a spline, and is derived
from both external and SNLS SN~Ia data as follows. We require that at
least one of the filters lie within 250\AA\ of rest frame $B$ in order
to ensure that $s_B$ is well determined.  However, in addition to
fits to individual filters, we also consider nearest neighbor combinations
of the SNLS filters ($\gmeg + \rmeg$, $\rmeg + \imeg$, and $\imeg + \zmeg$),
and include the SN if the mean wavelength of these filters meets the
above requirement.  In these cases $B$ is well bracketed, so $s_B$ is
also well constrained.  The stretch and (mean) rest-frame effective 
wavelength is recorded for each fit.  The results are shown in
figure~\ref{fig:stretchratio_sB}.

Alhough the relationship between $s_B$ and $s_j$ varies slightly with
wavelength, a general trend is apparent: $s_B$ is typically smaller
than $s_j$ for small $s_B$ and larger for large $s_B$. It is also
apparent that the low-z and SNLS samples probe different ranges in
$s_B$ (with the low-z sample containing more low-$s_B$ SNe and the
SNLS sample more high-$s_B$ SNe).  This is not surprising given
expected selection effects and the possibility of an evolving
population \citep[e.g.,][]{2007ApJ...667L..37H}.

\begin{figure*}
\plotone{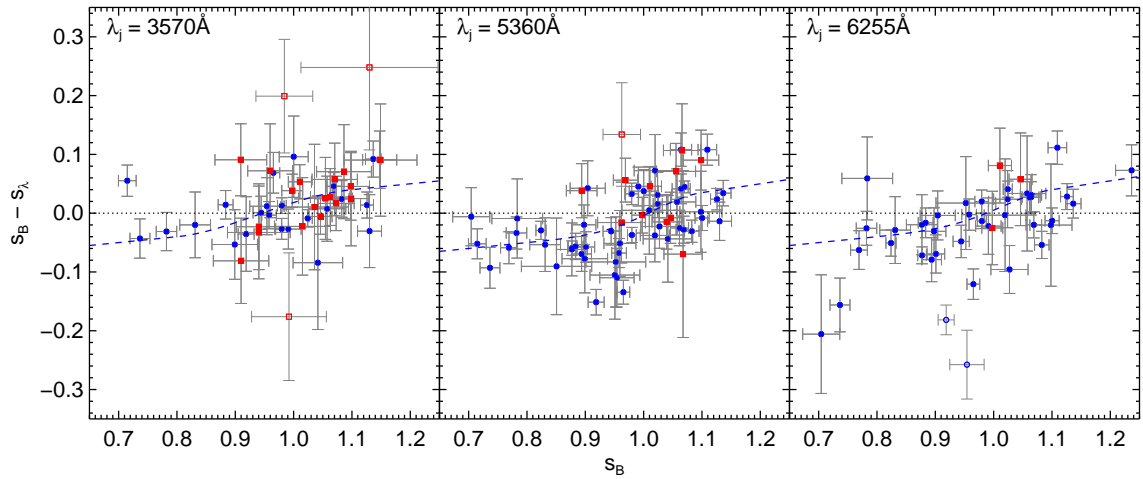}
\caption{Same as Fig.\ref{fig:stretchratio_sB}, but broken into bins in
  wavelength. Three example bins corresponding to $U$, $V$ and $R$
  are shown. The dashed line shows the best fitting
  function.\label{fig:stretch_wave}}
\end{figure*}

\begin{figure*}
\plotone{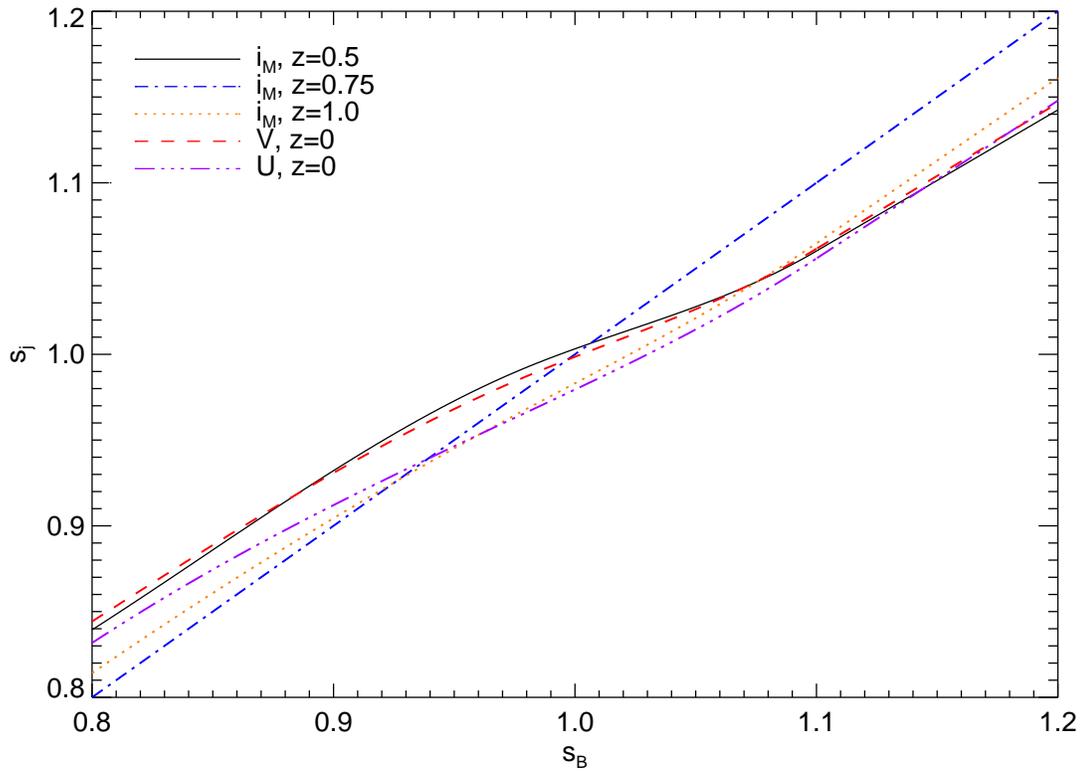}
\caption[Stretch function for various filters and $s_B$ values]
{Stretch in a particular observed filter ($s_j$) as a function
 of $s_B$ for various effective observed frame wavelengths $\lambda_j$.
 The MegaCam filter \imeg\ is a good match to $B$ at $z=0.75$, $U$ at $z=1$,
 and is between $B$ and $V$ at $z=0.5$. \label{fig:stretchfunction} }
\end{figure*}

To model the trends, we fit the relation between $s_B$ and $s_j$ using
least-squares in $\sim 100-200$\AA\ wide bins, using a cubic spline with
three knot points in $s_B$ smoothly joined to a linear relation at low
and high $s_B$.  Each wavelength bin contains 20--60 datapoints; three
example bins corresponding to $U$, $V$ and $R$ are shown in
figure~\ref{fig:stretch_wave}. The relationship between $s_B$ and $s_j$
at any intermediate $\lambda^j_{\mathrm{eff}}$ can be found by
linearly interpolating the fit coefficients in wavelength. We do not
probe all areas of $\lambda_{\mathrm{eff}}$ equally well, with
clusters around the low-z filter mean-wavelengths; this is an area
where larger samples of SNe~Ia will be extremely beneficial.  Examples
of the derived function are shown in figure~\ref{fig:stretchfunction}.

We can test the necessity of this more complex treatment of stretch by
applying the SiFTO model to the photometry from A06 (see 
\S \ref{subsec:hizcomparison} for more details).  For the
model described above, we find a \chisq\ of 2890 for 2822 degrees of
freedom.  If we remove the wavelength dependence of the stretch model
the \chisq\ climbs to 3037, and if we instead use a purely linear
model in place of the splines, the \chisq\ is 2935.  These correspond
to a decrease in likelihood by a factor of 70 and 3, respectively.
However, we note that the cosmological results and the scatter around
the best fit are almost identical in all three cases ( $< 0.001$ mag
difference in the RMS).  The lack of improvement in the relative
distances can be interpreted as an illustration of the difference
between a light-curve fitter and a distance estimator.

\subsection{Constructing the SED model}
\label{subsec:sedmodel}
\nobreak 
The starting point for our base SED model $\phi_0$ is an updated
version of the time series of spectral templates of H07 which
incorporates the additional low-z spectroscopy presented in
\citet{2008arXiv0803.1705M}.  The procedures of H07 are designed so
that the relative strengths of adjacent features are correct, but are
not intended to adjust the broadband colors or their evolution with
time, so some modification is necessary.  We developed a set of
light-curve templates that are used to address this issue.  These
templates are purely an intermediate step, and play no role in the
fitting process once $\phi_0$ is determined.

There are many approaches one might use to build the templates.  For
SiFTO, we construct a set of synthetic filters and then use SN data to
build light-curve templates in each of these filters.  We then adjust
$\phi_0$ to match these templates using interpolating splines as
described earlier.  This process requires $K$-correcting the observed
data points, which are at a variety of redshifts, to the nearest
template filter.  Since the $K$-correction depends on the model SED,
this procedure is carried out iteratively. We follow a fairly standard
approach by parameterizing each light-curve template as a set of cubic
splines with the knot points tuned by hand to best describe the data
(e.g., \citet{2001ApJ...558..359G}).  The number of knots ranges from
3 (for poorly constrained filters) to 8 for the redder filters which
have a more complex template.  Since cubic splines are used, the
templates are only smooth up to first order in the derivatives;
applications which require higher degrees of smoothness should not use
SiFTO.  The splines are joined smoothly to a polynomial at early
times, and an exponential decay function at late times.  The location
of the join can be set for each filter; generally the early time
polynomial joins the splines around epoch $-10$, and the exponential
tail around $+50$, although in some filters this is much lower due to
the lack of good late-time data.  The early-time polynomial is of the
form $f = a_p \left( t - t_{\mathrm{exp}} \right)^2 + b_p \left( t -
t_{\mathrm{exp}} \right)^3$, where $t_{\mathrm{exp}}$ is the date of
zero flux (i.e., the date of explosion).  Therefore, for our template,
the colors of the SN at early times are not constant in all cases due
to the cubic term.  We require the polynomial coefficients to be
positive to ensure that the early time light-curve increases
monotonically at early times, which we expect on physical grounds.
Linear and quartic terms were not found to improve the fits.

The choice of which filters to construct templates in is fairly
arbitrary.  If only low-z data is used for training, then a natural
choice would be the Landolt $UBVRI$ filters, redshifted slightly to
the mean redshift of the sample.  Since we use both low-z and SNLS
data, this is no longer optimal.  We want to make use of information
blue-ward of $U$, and it is also useful to insert filters between the
more widely spaced Landolt filters to make better use of our
observations and improve the smoothness of the model.

\begin{figure*}
\plotone{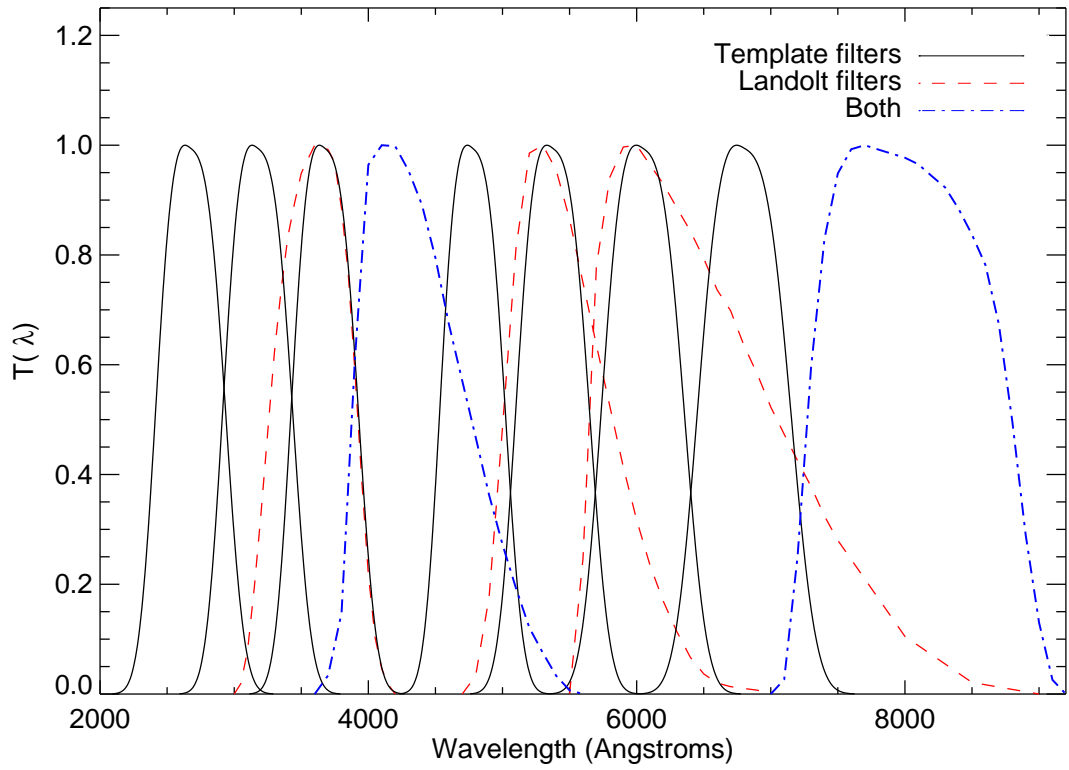}
\caption[Filters used when constructing the SED template]
{Filters used when constructing the SED template $\phi_0$
 (solid lines) compared with $UBVRI$ (dashed lines).  The
 $B$ and $I$ filters (dot-dashed lines) are shared by both.
 The normalization of the filter responses is arbitrary.
 \label{fig:templatefilters} }
\end{figure*}

We use a set of synthetic logarithmically spaced filters between 4800
\AA\ and 6800 \AA , and also include $B$ and $I$.  In addition, we
include three bluer filters which are blueshifted and widened
versions of the first log filter.  The $B$ filter is included to make
it easier to compare the resulting template with previous work, and
$I$ to ensure well-behaved boundary conditions while building
$\phi_0$.  We do not use the part of the model redder than 7100 \AA\
when applying SiFTO to observations because our stretch
parameterization is not expected to work well.  The synthetic log
filters are denoted as LOG5 -- LOG8, and the three bluer filters are
ARB0 -- ARB2.  The light-curves in ARB0 and ARB1 are determined purely
by SNLS data.  Our filters are shown in relation to the Landolt
filters in figure~\ref{fig:templatefilters}.  ARB2 matches $U$ (but is
slightly redder), LOG6 roughly matches $V$, $R$ is split up into two
narrower filters which are better matches to the SNLS observations
(LOG7 and LOG8), and there is an additional filter between $B$ and $V$
(LOG5).  The filters are summarized in table~\ref{tbl:templateparams}.
The SiFTO model should only be used for observations from 2700 to 7100
\AA\ (rest frame).

We use cubic terms in the early-time polynomials for the ARB0, ARB2,
and LOG5 filters.  For ARB0, including this term reduces the \chisq\
of the template fit to our training sample from 171 for 155 degrees of
freedom to 167.  For ARB2 the \chisq\ goes from 4613 to 4515 for 1605
degrees of freedom, and for LOG5 from 2280 for 1004 degrees of freedom
to 2001.  We caution that these \chisq\ values do not incorporate the
template error snake as described in \S \ref{subsec:errorsnake}, and
hence we do not expect the reduced \chisq\ to be close to unity.  It
is difficult to evaluate how significant the improvements in the
\chisq\ are, since the process of building the error snake involves
inflating the errors to take into account the residuals from the
model.  However, in the other filters the best fit value of the cubic
coefficient is zero.

The procedure of building the templates in each filter is as follows:
First, all of the data is fit using SiFTO with the current version of
$\phi_0$.  The observed data is $K$- and MW-extinction corrected to
the best matching template filter, and the timescale is normalized
using $t_0$ and the stretch in each filter $s_j$.  The data
points from each SN in each filter are normalized to an arbitrary
value in order to remove the colors of the individual SNe from the
template.  The templates are then fit to the datapoints in each
filter, and $\phi_0$ is adjusted using the spline procedure to match
the model.  The procedure is then iterated until convergence, which
typically requires $3-5$ iterations.  We also adjust the template so
that $t = t_0$ represents the peak flux in $B$, and adjust $s_B$ so
that the mean over our training sample is unity.  

It is convenient for our template to have roughly the colors of a
fiducial SN~Ia.  Given the asymmetry of the color distribution we use
the median color to define these values.  There is no single SN with
exactly this color, and if there were it would not constrain all of
our template filters.  Therefore, we take the mean 10\% of the SN
closest to the median color to define each color, and use this in the
normalization process above.  Note that this process does not affect
the resulting light-curve fits as long as sharp, unphysical gradients
are avoided.  Changing the colors individually by up to 0.3 mag has no
effect on the resulting fits because our procedure adjusts the SED
template to match the observed colors.

With the above ingredients, we can write $\phi$ as
\begin{multline}
 \phi \left( \lambda, t, t_0, n_i, s_B \right) =
    \phi_0 \left(\lambda ,\frac{t - t_0}{s_B \left(1+\zhel\right)} \right) \\
     \times {\mathcal S} \left( \lambda ; {\mathcal R}_i 
       \left( t, t_0, s_B, s_j, \zhel \right) n_i \right)
\end{multline}
where $\mathcal S$ is the spline used to make the SED have the desired
colors, and not that used to construct $\phi_0$.  $\mathcal
S$ is uniquely determined by the combination of the $\mathcal{R}_i
n_i$ products, $\phi_0$, and the observed filters.  $\phi$ can then be
used to evaluate the flux in any desired filter, and not just those
actually observed.

\begin{deluxetable}{lcrrrrr}
\tablecaption{Template Spline Parameters}
\tablehead{
 \colhead{Filter} & \colhead{ $\left< \lambda \right>$ } &
 \colhead{ $\lambda_{\mathrm{eff}}$ \tablenotemark{a} } &
 \colhead{ FWHM } & \colhead{ $t_{\mathrm{poly}}$ \tablenotemark{b} } & 
 \colhead{ $N_s$ \tablenotemark{c} } &
 \colhead{ $t_{\mathrm{tail}}$ \tablenotemark{d} }
}
\startdata
ARB0 & 2689 & 2798 & 529  & $-11.5$ & 3 & 12 \\
ARB1 & 3187 & 3206 & 530  & $-11$   & 4 & 25 \\
ARB2 & 3685 & 3704 & 510  & $-11$   & 5 & 30 \\
B    & 4413 & 4330 & 893  & $-10$   & 6 & 50 \\
LOG5 & 4784 & 4755 & 497  & $-10.5$ & 5 & 41 \\
LOG6 & 5382 & 5363 & 559  & $-10$   & 6 & 48 \\
LOG7 & 6055 & 6030 & 628  & $-10$   & 6 & 50 \\
LOG8 & 6812 & 6854 & 707  & $-11$   & 7 & 47 \\
I    & 8060 & 8008 & 1526 & $-11$   & 7 & 47 \\
\enddata
\tablecomments{Information about the template filters and the
 functional form used to fit each filter when deriving $\phi_0$.
 Except for $B$ and $I$, these filters are synthetic.
 These values are essentially hand-tuned. \label{tbl:templateparams}
}
\tablenotetext{a}{The effective wavelength for this filter when observing
  a fiducial SN~Ia at $z=0$.}
\tablenotetext{b}{The epoch of the join between the splines and the 
  polynomial early form, relative to the epoch of $B$ maximum.}
\tablenotetext{c}{The number of internal knots in the spline between
 $t_{\mathrm{poly}}$ and $t_{\mathrm{tail}}$}
\tablenotetext{d}{The epoch of the join between the splines and the 
  exponential late form, relative to the epoch of $B$ maximum.}
\end{deluxetable}

\subsection{The template error snake}
\label{subsec:errorsnake}
\nobreak 
SiFTO incorporates an estimate of the uncertainty in $\phi_0$.  This
represents both the uncertainty in our modeling process, and, much
more importantly, some measure of the intrinsic variability of SNe~Ia.
SALT2 and MLCS2k2 also incorporate an error snake, but some older
fitters do not.  The uncertainty in the template is very important
when fitting low-z data, where the photometric uncertainties can often
be quite small.  It also plays a critical role when evaluating \gmeg\
and \rmeg\ observations of high-z SN, where the model is more poorly
constrained.  We tabulate the template error at the wavelengths of
each of the template filters in one day bins and then linearly
interpolate between these errors in wavelength space to get the error
in the SED, which gives us the $\sigma_F$ term in
equation~\ref{eq:lcfit}.

There are several methods used to derive the error snake, which give
similar results.  The most direct approach is to fit all of the SNe
using SiFTO, and then determine the additional error that would make
the $\chisq$ of the data relative to the model be 1 per degree of
freedom.  The difficulty with this approach is that at some epochs the
measurement errors of individual observations dominate the scatter.  The
SNLS error bars are roughly constant with epoch in flux space because
it is a rolling search, but a 10\% variation in the template flux at
any given epoch is much larger in absolute flux at peak than at late
or early times.  Therefore, measuring the same relative variation at
late times is more difficult, requiring an accurate cancellation of
two numbers subtracted in quadrature.  Compensating for this fact is
the tendency of the template error to be larger in relative terms at
late times which partially arises from the fact that we have
renormalized the fits to each SN at peak.  In order to improve our
sensitivity at late and early epochs, we adopt a slightly modified
procedure.  In the core of the light curve ($-10$ to $+15$ days) we
use the \chisq\ method directly, but at late and early times we
substitute a measure based on the RMS, but multiplied by the ratio
between the \chisq\ result and the RMS as measured in the core.  The
results of this procedure are generally quite similar to that of the
raw \chisq\ method (within about $10\%$).

\begin{figure*}
\plotone{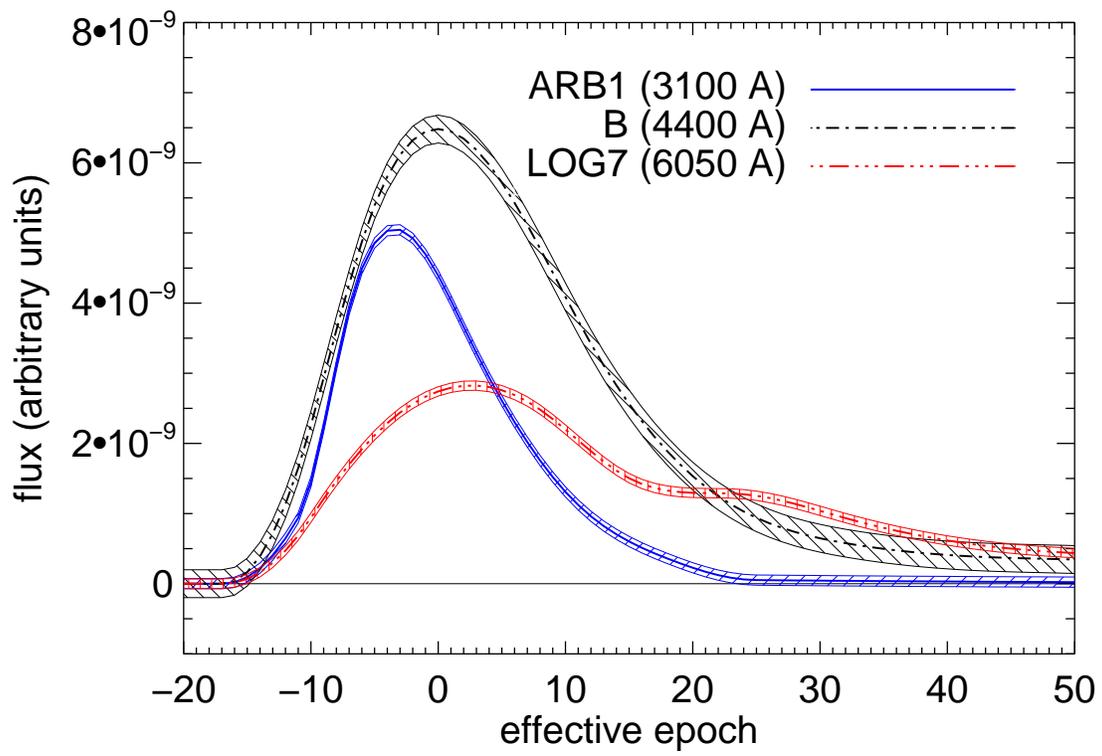}
\caption[Sample error snakes for some of the template filters]
{ Example error snakes for some of the template filters.  ARB1
 is bluer than $U$ and LOG7 is the bluer of the two filters
 that overlap with $R$. Both ARB1 and LOG7 are synthetic filters.
 \label{fig:errorsnake} }
\end{figure*}

The other method we consider is to use bootstrap-with-replacement to
estimate the template errors.  This gives similar results, but the
\chisq\ + RMS approach is more robust and slightly more conservative
in that it gives larger errors (by $\sim 20\%$ on average), so we use
this for our final model.  The error snake is used iteratively using the
template construction process.  Some examples are shown in
figure~\ref{fig:errorsnake}.

Since we usually have a large number of observations in each epoch bin
for each filter, the formal statistical error in how well we can
measure the mean template is much smaller ($\sim 100$ times) than the
template error we quote.  Some of this is probably due to failures in
the photometry, uncertainties in the $K$-corrections, and inconsistencies
between different samples.  However, looking at any of these in detail
(for example, by considering only data from a single source such as SNLS,
or looking at SN in a narrow redshift range where $K$-correction uncertainties
are less important in a relative sense), none of these effects is the right
order of magnitude to explain the observed scatter.  We conclude that
SN variability is the dominant factor in the template error, except in
our bluest artificial filter where the sample is small.

Like SALT2, our error snake does not incorporate off diagonal terms --
i.e., the residual from the model at a given epoch is assumed to be
uncorrelated with nearby epochs.  Given the complexity of our model
space, the current data set does not provide robust constraints on
these terms.  MLCS2k2, which has a simpler model and does not work
with a SED, includes off-diagonal terms with simplifying assumptions.

\subsection{Modeling systematics using SiFTO}
\label{subsec:systematics}
\nobreak 
Modeling of systematic errors is becoming increasingly important for
SN cosmology.  One issue which has been neglected is that because our
SN model is empirically derived from SN data, the effects of the
systematics on the model must also be considered in the error budget.
This is important even if no high-z data is used in the training.  An
advantage that SiFTO has over most other light-curve fitters is that
the effects of epoch-independent systematics can be studied simply by
re-deriving the color relation, rather than having to fully retrain
all aspects of the model.  For example, a change in the SNLS \gmeg\
zeropoint data does not affect the SED model because during the
training process the data from each SN is normalized to an arbitrary
level.  It will, however, affect the relation between the measured
colors of the SN population.  Approaches which fold the modeling of SN
colors into the model derivation more directly (SALT/SALT2, MLCS2k2)
do not have this luxury.

This does not apply to any effect which can change the
shape of the light curve, such as uncertainties in the filter
responses.  However, in most cases these are second order effects
compared with the changes in the derived color relation.  This makes
systematic analysis with SiFTO comparatively easy.

\section{DATA}
\label{sec:data}
\nobreak
Our training sample consists of photometry for nearby SNe~Ia from the
literature as well as high-redshift SNLS observations.  The low-z data
generally has higher signal-to-noise ratios, but are also much more
heterogeneous.  More importantly, the SNLS data offers considerably better
wavelength coverage than the low-z sample, both in the near-UV and 
between the standard Landolt filters.  Furthermore, the rest-frame $U$
band data from SNLS is better calibrated and more reliable than that
from the low-z data.

The photometric references of the 72 low-z SNe used in this paper are
given in table~\ref{tbl:lowzsample}.  We have not included sub-luminous
(SN 1991bg-like) SNe~Ia in our sample because they are not well
represented by our base SED, and have excluded extremely peculiar SNe
such as SN 2002cx \citep{Li2002cx:03}.  Furthermore, we have not
included all of the available photometry for every SN.  In cases where
photometry is available from multiple sources, we compare the
light-curves from multiple bands and, if they are not consistent,
choose some subset of the sources.  We try to choose the photometry
that has the best coverage, or that from the data source with the most
other SNe (i.e., we prefer observations from a large sample like
\citet{2006AJ....131..527J}).  We also exclude individual bands on a
SN by SN basis where the photometry seems to be internally
inconsistent or has additional problems (e.g., the $U$ band
data for SN 1999ee).

In addition we use 98 high-redshift SNLS SNe from the first three
years of the survey.  Our training process requires that the peak of
each observed filter be fairly well constrained so that we can
re-normalize the light-curves accurately, a necessity since we do not
impose a color relation during the training process or fits.
Therefore, to use an individual bandpass for a SN, we require that it
have at least one datapoint within 5 rest-frame days of maximum.  This
requirement is far more stringent than is necessary to carry out a
cosmological analysis, so less than half of our high-z data are used in
the training process.  We use only SNe with firm spectroscopic
identification.  The H07 template makes use of SNLS spectra, which are
particularly critical when constraining the near-UV SED.  As was the
case for the low-z sample, we exclude SNe which are known to be unusual,
such as the super-Chandrasekhar mass 03D3bb
\citep{2006Natur.443..308H}.

For SNLS, we adopt the radially-dependent SAGEM filter scans for the
MegaCam filters, corrected for the f/4 converging beam and
incorporating the Mauna Kea extinction curve at the mean survey
airmass of 1.2.  A06 used radially averaged versions of these scans.
For the low-z data we adopt the \citet{1990PASP..102.1181B}
realization of the Landolt filter functions.  Ideally we would have
natural system magnitudes and bandpasses for all of the low-z
photometry, but this information is generally not available.  Hopefully
this will not be the case for future low-z data sets. 
Currently we do not implement any airmass dependence in our assumed
filter functions, since tests show that this is negligible in
comparison with the other bandpass uncertainties.

\begin{deluxetable}{lll}
\tablecaption{Low-Redshift SNe~Ia Used to Build the Rest-Frame SED Model}
\tablehead{
 \colhead{Name} & \colhead{ Filters } & \colhead{ Source \tablenotemark{a} }
}
\startdata
1981D  & $UBV$    & H91 \\
1986G  & $UBV$	  & P87 \\
1989B  & $UBV$	  & W94 \\
1990N  & $UBV$	  & L98 \\
1990O  & $BV$	  & H96 \\
1990af & $BV$	  & H96 \\
1992ae & $BV$	  & H96 \\
1992al & $BVRI$	  & H96 \\
1992bc & $BVRI$	  & H96 \\
1992bg & $BVI$	  & H96 \\
1992bh & $BVI$	  & H96 \\
1992bl & $BV$	  & H96 \\
1992bo & $BVRI$	  & H96 \\
1992bp & $BVI$	  & H96 \\
1992br & $BV$	  & H96 \\
1992bs & $BV$	  & H96 \\
1993B  & $BV$	  & H96 \\
1993H  & $BVRI$	  & H96, A04 \\
1993O  & $BVI$	  & H96 \\
1993ag & $BV$	  & H96 \\
1994M  & $BVRI$	  & R99, A04 \\
1994S  & $BVI$	  & R99 \\
1994ae & $BVRI$	  & R05, A04 \\
1995D  & $BVRI$	  & R99 \\
1995E  & $BVRI$	  & R99 \\
1995ac & $BVRI$	  & R99 \\
1995al & $BVRI$	  & R99 \\
1995bd & $BVRI$	  & R99, A04 \\
1996C  & $BVRI$	  & R99 \\
1996X  & $UBVRI$  & R99, S01 \\
1996ab & $BV$	  & R99 \\
1996bl & $BVRI$	  & R99 \\
1996bo & $BVRI$	  & R99 \\
1997E  & $UBVRI$  & J06 \\
1997bp & $UBVRI$  & J06 \\
1997dg & $UBVRI$  & J06 \\
1997do & $UBVR$	  & J06 \\
1998V  & $UBVRI$  & J06 \\
1998aq & $UBVRI$  & R05 \\
1998bu & $UBVRI$  & J99, S99 \\
1998dx & $UBV$	  & J06 \\
1998es & $UBVRI$  & J06 \\
1999aa & $UBVRI$  & J06, A04, K00 \\
1999ac & $UBVRI$  & J06 \\
1999aw & $BVRI$	  & S02a \\
1999cc & $UBVR$	  & J06, K06 \\
1999cl & $BV$	  & J06, K06 \\
1999dk & $UBVI$	  & K01, A04 \\
1999dq & $UBVRI$  & J06 \\
1999ee & $BVRI$	  & S02b \\
1999ek & $BVRI$	  & J06, K04 \\
1999gd & $UBVR$	  & J06 \\
1999gp & $UBVI$	  & J06, K01 \\
2000ca & $UBVRI$  & K04 \\
2000cn & $UBVRI$  & J06 \\
2000dk & $UBVI$	  & J06 \\
2000fa & $UBVRI$  & J06 \\
2001V  & $UBVR$	  & V03, L06a \\
2001ba & $BVI$	  & K04 \\
2001bt & $BVRI$	  & K04 \\
2001cn & $BVRI$	  & K04 \\
2001cz & $BVRI$	  & K04 \\
2001el & $BVRI$	  & K03 \\
2002bf & $BVRI$	  & L05 \\
2002bo & $BVRI$	  & K04 \\
2002er & $UBVR$	  & P04 \\
2003cg & $UBVRI$  & ER06 \\
2004S  & $UBVRI$  & K07 \\
2004eo & $UBVRI$  & P07b \\
2004fu & $BV$	  & T06 \\
2005am & $BVR$	  & L06a \\
2005cf & $UBVRI$  & P07a \\
\enddata
\tablenotetext{a}{ Photometry References: H91 \citet{1991AJ....102..208H},
 P87:  \citet{1987PASP...99..592P}, W94:  \citet{1994AJ....108.2233W},
 H96:  \citet{1996AJ....112.2408H}, L98:  \citet{1998AJ....115..234L}, 
 J99:  \citet{1999ApJS..125...73J}, R99:  \citet{1999AJ....117..707R}, 
 S99:  \citet{1999AJ....117.1175S}, K00:  \citet{2000ApJ...539..658K},
 K01:  \citet{2001AJ....122.1616K}, S01:  \citet{2001MNRAS.321..254S}, 
 S02a: \citet{2002AJ....124.2905S}, S02b: \citet{2002AJ....124.2100S},
 K03:  \citet{2003AJ....125..166K}, V03:  \citet{2003AaA...397..115V},
 A04:  \citet{2004MNRAS.349.1344A}, K04:  \citet{2004AJ....128.3034K}, 
 P04:  \citet{2004MNRAS.355..178P}, L05:  \citet{2005ApJ...632..450L},
 R05:  \citet{2005ApJ...627..579R}, ER06: \citet{2006MNRAS.369.1880E},
 K06:  \citet{2006AJ....131.1639K}, J06:  \citet{2006AJ....131..527J}, 
 L06a: \citet{2006AJ....132.2024L}, L06b: \citet{2006PASP..118...37L},
 T06:  \citet{2006astro.ph..6051T}, K07:  \citet{2007AJ....133...58K},
 P07a: \citet{2007MNRAS.376.1301P}, P07b: \citet{2007MNRAS.tmp..321P}
} 
\label{tbl:lowzsample}
\end{deluxetable}

\section{RESULTS OF THE TRAINING}
\label{sec:results}
\begin{figure*}
\plotone{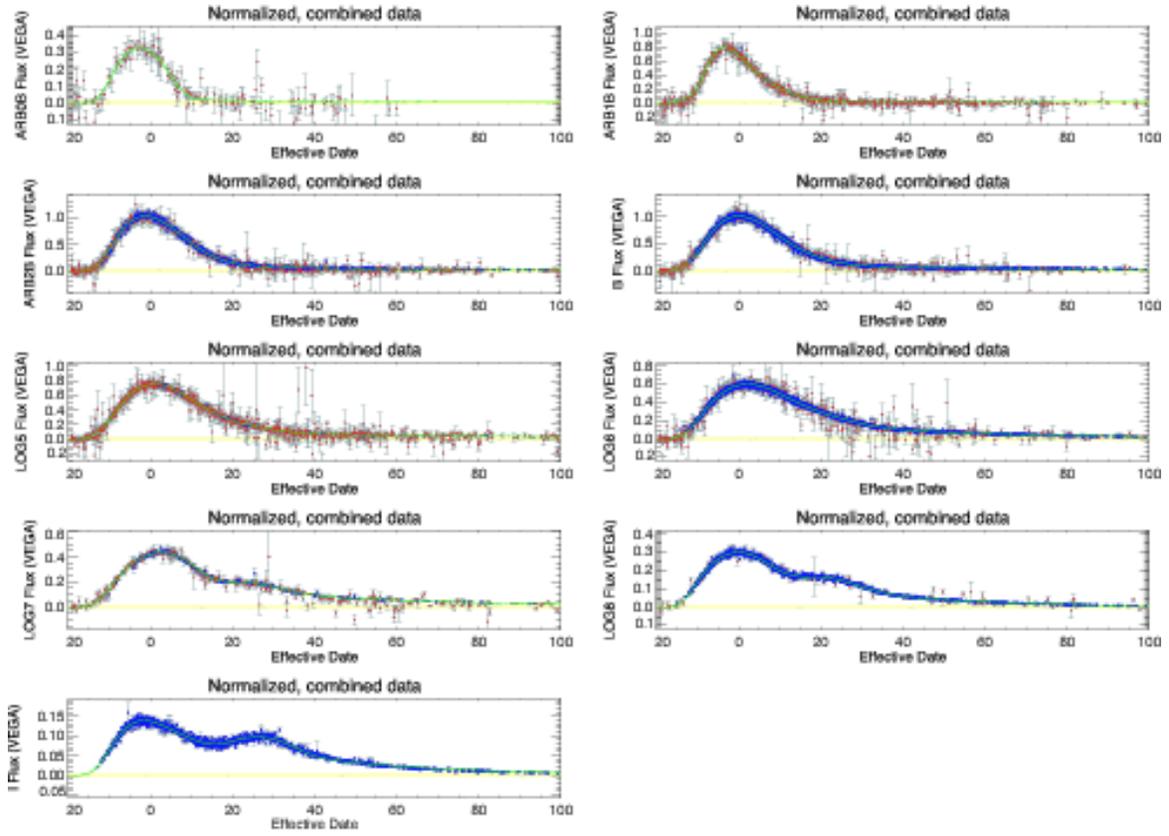}
\caption[Light-curve template fits]{Light-curve template fits
 in our filters. The red squares are SNLS data, and the blue circles
 are low-z data.  The fluxes have been normalized so that the $B$
 template has a peak value of unity, and the colors to be representative
 of a typical SN~Ia.  A full resolution version is available at
 {http://qold.astro.utoronto.ca/conley/sifto\_fullres.pdf}.
\label{fig:templatefits} }
\end{figure*}

\begin{figure*}
\plotone{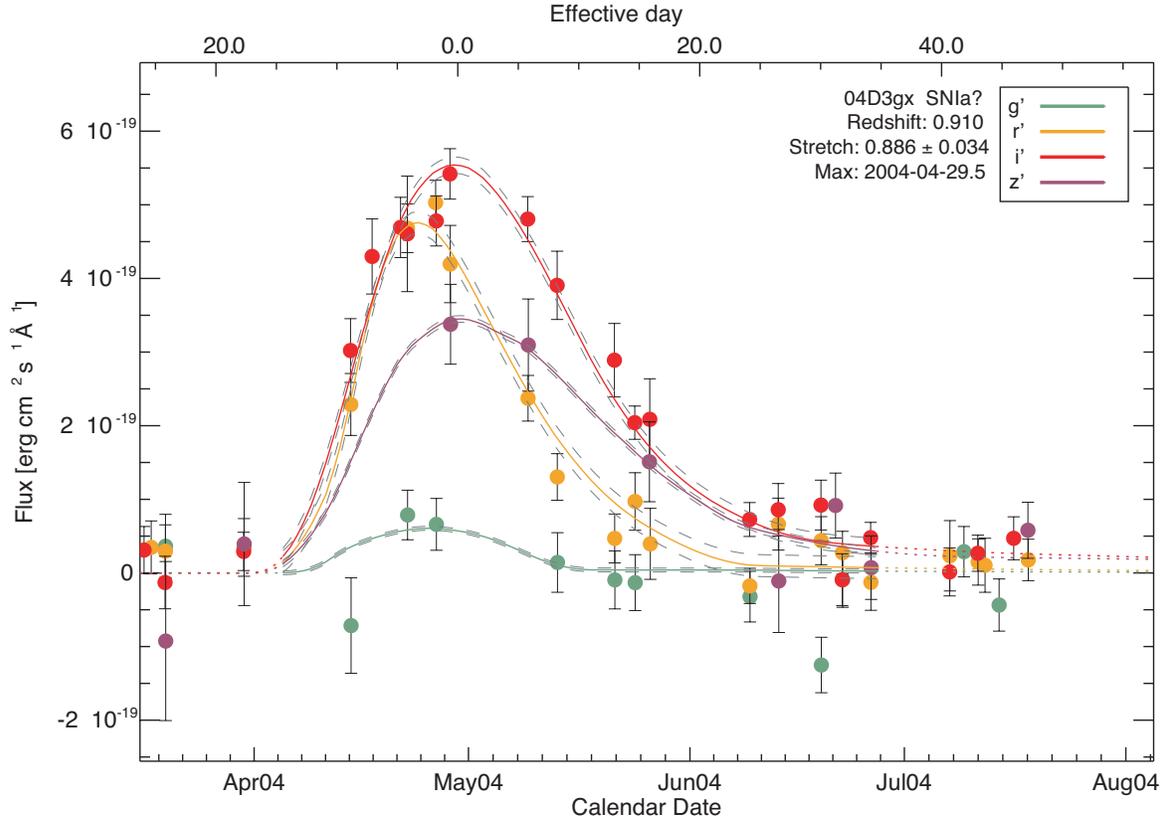}
\caption{Sample SiFTO fit to 04D3gx, a high-redshift ($z=0.91$) SNLS
 SN~Ia.  The dashed lines are the error snake for the template, and
 do not include the uncertainties in the light-curve
 parameters.  This SN is from the first-year sample, and therefore
 does not represent the quality of light-curves currently obtained by
 SNLS.  In particular, the signal-to-noise ratio in \zmeg\ has improved
 considerably in the third-year sample due to longer exposure times.
 \label{fig:example_one} }
\end{figure*}

\begin{figure*}
\plotone{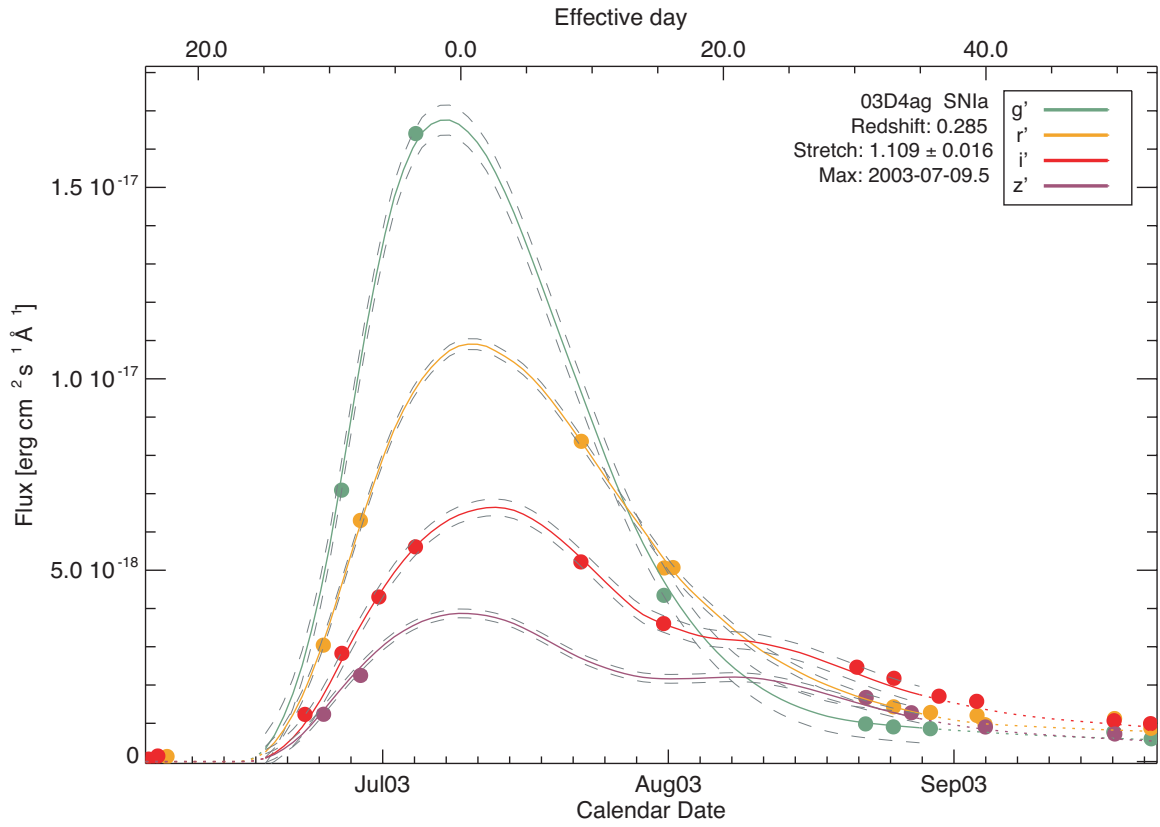}
\caption{Sample SiFTO fit to 03D4ag, a low-redshift ($z=0.285$) SNLS SNe~Ia.
 The dashed lines are the error snake for the template, and do not
 include the uncertainties in the light-curve parameters.
 \label{fig:example_two} }
\end{figure*}

\nobreak 
The light-curve templates can be seen in
figure~\ref{fig:templatefits}.  The different manner in which the SNLS
and low-z data sets sample wavelength space is clear -- only the SNLS
data offer any constraints blue-ward of $U$ (ARB0 and ARB1), and the
wavelength region between $B$ and $V$ (LOG5) and $V$ and $R$ (LOG7),
while the nearby data provide most of the constraints in $RI$.
However, in the filters where substantial coverage is provided by both
sets (LOG3, $B$, LOG6), the agreement is impressive.  In addition, the
homogeneity of SNe~Ia light curves is also apparent.  Note that in the
$I$ band extensive culling has been performed because many SN do not
match the template.  This is acceptable because we are only including
$I$ to ensure well behaved boundary conditions, and do not use this
part of the model directly when applying the model to SN observations.
Example fits to high and low redshift SNLS SNe are shown in
figures~\ref{fig:example_one} and \ref{fig:example_two} (the same
SN were used as examples in G07).

\begin{figure*}
\plotone{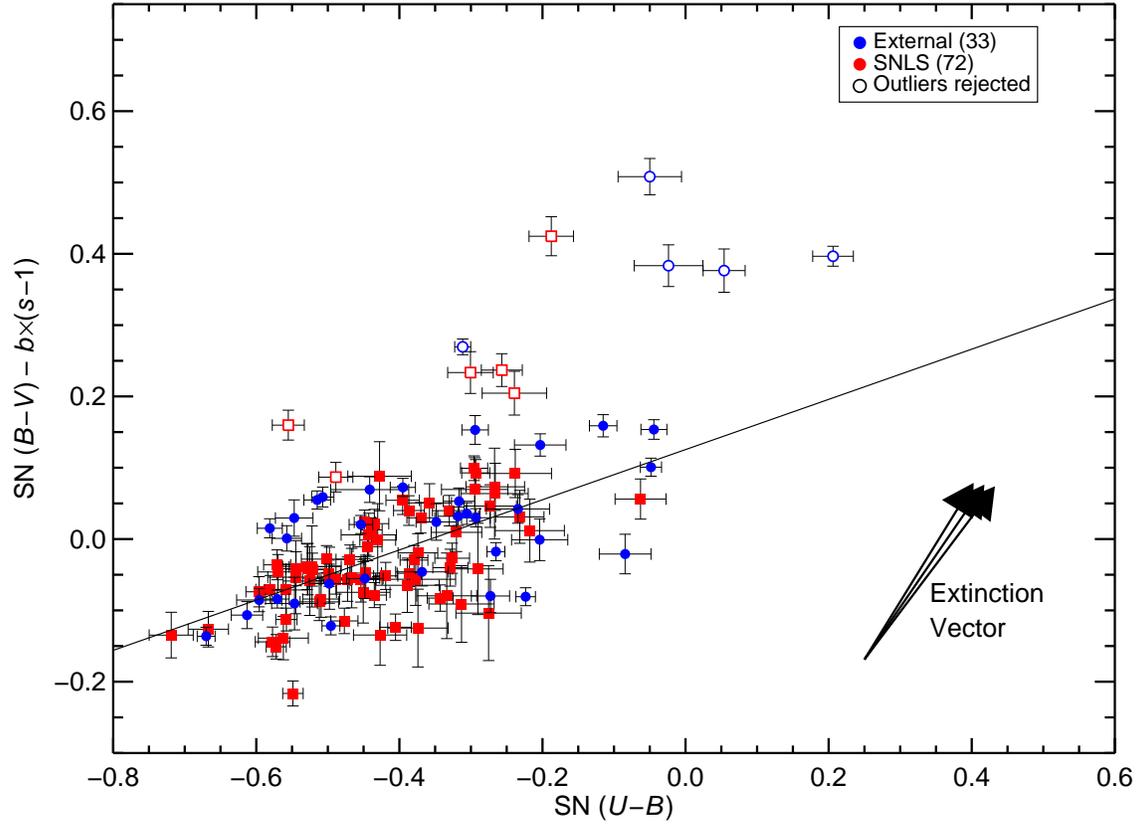}
\caption[$U-B$ vs.\ $B-V$ relation]
{Derived $U-B$ vs.\ $B-V$ relation using low-z (blue circles)
 and SNLS data (red squares).  The best fit relation to the full sample
 is shown as a solid line.  Also shown is the
 relation one would expect from Milky-Way like dust with 
 $R_V = \left[1.6,3.1,4.6\right]$. The points rejected as outliers
 are open. \label{fig:colorparams} }
\end{figure*}

\begin{figure*}
\plotone{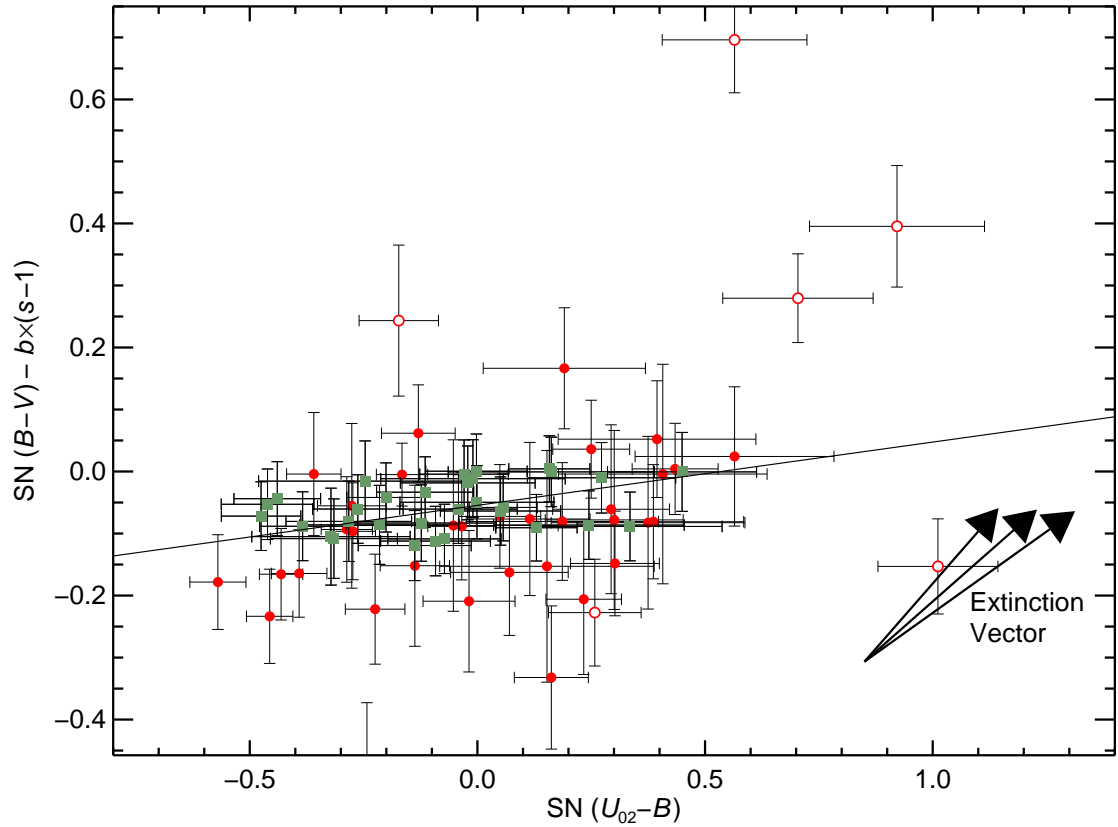}
\caption[$U_{02}-B$ vs.\ $B-V$ relation]
{Derived $U_{02}-B$ vs.\ $B-V$ relation using SNLS data, similar
 to figure~\ref{fig:colorparams}.  The red circles are data for which
 $U_{02}-B$ and $B-V$ were measured, the green squares are where
 $U-B$ was used to predict $B-V$. \label{fig:colorparams02} }
\end{figure*}

The parameters of the color relations used to determine parameter
\scriptc\ (equation~\ref{eq:colorrelation}) are given in
tables~\ref{tbl:colorparams} and \ref{tbl:colorparams02}, and the
$U-B$ and $U_{02}-B$ relations shown in figures~\ref{fig:colorparams}
and \ref{fig:colorparams02}.  The $U-B$ relation is derived from 72
SNLS and 33 low-z SNe, and the $U_{02}-B$ relation from 63 SNLS SNe.
Here, because there are few SNe with both $V$ and $U_{02}$
measurements, we include data with no rest-frame $V$ observations by
using their $U-B$ colors to predict $B-V$, incorporating the
additional uncertainty due to $\sigma_{\mathrm{clr}}$ and the
appropriate covariances.  The requirements to be included in
the color relation fit are less stringent than those required to be
used when measuring $\phi_0$ since we do not require that the fits to
each filter be strongly constrained in isolation.  The primary
limitation is instead that most SNe do not have simultaneous
measurements of $U-B$ and $B-V$.  Carrying out this test with SALT2
gives similar values for the $\sigma_{\mathrm{clr}}$ parameters.  It
is interesting to consider the residuals from the $U-B$ versus $B-V$
relation.  The slope of the dominant relation is clearly inconsistent
with that expected from extinction.  However, there are a number of
outliers that are displaced in a direction that could be explained by
extinction.  This leads us to hypothesize that the color relations for
the majority of SNe in our sample are dominated by intrinsic effects,
with a minority instead dominated by dust.

Following G07, we can get some feel for how useful extending the model
blue-ward of $U$ is be comparing measurements of \scriptc\ if the
model is cut at different wavelengths.  This is shown in
table~\ref{tbl:uveffects} for 04D3cy ($z=0.64$) and 04D3gx ($z=0.91$).
The errors include statistical uncertainties, template errors,
and $K$-correction errors as described in H07.  The benefits for the
higher redshift SN are clear.

\begin{deluxetable}{lrr}
\tablecaption{SiFTO Color Relation Parameters}
\tablehead{
  \colhead{Parameter} & \colhead{Value} & \colhead{Error}
}
\startdata
\cutinhead{SNLS + low-z}
$a$                      &  0.352 & 0.049 \\
$b$                      &  0.218 & 0.064 \\
$c$                      & $-$0.051 & 0.007 \\
$\sigma_{\mathrm{clr}}$  &  0.055 & 0.005 \\
\cutinhead{SNLS only}
$a$                      &  0.435 & 0.067 \\
$b$                      &  0.302 & 0.076 \\
$c$                      & $-$0.067 & 0.008 \\
$\sigma_{\mathrm{clr}}$  &  0.044 & 0.006 \\
\cutinhead{low-z only}
$a$                      &  0.259 & 0.090 \\
$b$                      &  0.145 & 0.132 \\
$c$                      & $-$0.037 & 0.016 \\
$\sigma_{\mathrm{clr}}$  &  0.071 & 0.011 \\
\enddata
\tablecomments{ The coefficients for the $U-B$ vs.\ $B-V$
 color relation of equation~\ref{eq:colorrelation}.
 Only the statistical errors are given.  See 
 \S \ref{subsec:colorcompare} for a comparison
 of the SNLS and low-z values. \label{tbl:colorparams} }
\end{deluxetable}

\begin{deluxetable}{lrr}
\tablecaption{SiFTO Color Relation Parameters: $U_{02}-B$}
\tablehead{
  \colhead{Parameter} & \colhead{Value} & \colhead{Error}
}
\startdata
$a_{02}$                   &  0.121 & 0.050 \\
$b_{02}$                   &  0.172 & 0.138 \\
$c_{02}$                   & $-$0.110 & 0.027 \\
$\sigma_{02\mathrm{clr}}$  &  0.050 & 0.008 \\
\enddata
\tablecomments{ The coefficients for the $U_{02}-B$ vs.\ $B-V$
 color relation. \label{tbl:colorparams02} }
\end{deluxetable}

\begin{deluxetable}{llrlll}
\tablewidth{0pt}
\tabletypesize{\tiny}
\tablecaption{Effect of Including Near-UV Information}
\tablehead{ \colhead{$\lambda_{\mathrm{min}}$} & \colhead{bands} &
 \colhead{ \scriptc } & \colhead{ $w_{U02B}$ \tablenotemark{a} } &
 \colhead{ $w_{UB}$ \tablenotemark{b} } &
 \colhead{ $w_{BV}$ \tablenotemark{c} } }
\startdata
\cutinhead{SNLS-05D3lb $z=0.65$}
3980 \AA & \imeg \zmeg             & $0.066 \pm 0.097$ & 0    & 0    & 1    \\
3250 \AA & \rmeg \imeg \zmeg       & $0.008 \pm 0.040$ & 0    & 0.69 & 0.31 \\
2700 \AA & \gmeg \rmeg \imeg \zmeg & $-0.002 \pm 0.038$ & 0.43 & 0.40 & 0.18 \\
\cutinhead{SNLS-04D3gx $z=0.91$}
3980 \AA & \imeg \zmeg             & $-0.106 \pm 0.080$ & 0    & 1    & 0    \\
3250 \AA & \rmeg \imeg \zmeg       & $-0.100 \pm 0.061$ & 0    & 1    & 0    \\
2700 \AA & \gmeg \rmeg \imeg \zmeg & $-0.097 \pm 0.040$ & 0.60 & 0.40 & 0.00 \\
\enddata
\tablenotetext{a}{The weight given to $U_{02}-B$ when forming \scriptc .}
\tablenotetext{b}{The weight given to $U-B$ when forming \scriptc .}
\tablenotetext{c}{The weight given to $B-V$ when forming \scriptc .}
\tablecomments{ The errors in \scriptc\ for two SNLS SNe when the model is
 cut at different wavelengths.  The cuts correspond to including
 all of $U_{02}UBV$, removing $U_{02}$, and removing $U$.  
 Note that the rest-frame $V$ is not measured for SNLS-04D3gx.  The improvement
 in the measured color for 04D3gx when \rmeg\ is added is due to the
 improved constraints on the other light-curve parameters such as the 
 date of maximum, even though it is not formally used in the color. 
 \label{tbl:uveffects} }
\end{deluxetable}

The stretch is only defined relative to some fiducial template, and
does not have any independent meaning when comparing different
fitters, or even different versions of the same fitter.  Instead, a
variable like $\Delta m_{15}$ can be used which has a more physical
definition.\footnote{The quantity $\Delta m_{15}$ is the amount in
magnitudes which the $B$ band light curve has dimmed 15 rest frame
days after peak.}  However, in practice, because light-curve fits are
relatively insensitive to the exact shape of the light-curve template
near peak, even $\Delta m_{15}$ is somewhat ambiguous, and will change
when the light-curve model is updated.  Therefore, we simply provide
the relation between $\Delta m_{15}$ and $s_B$ for our template:
\begin{equation*}
 \Delta m_{15} = 1.00 -1.63 \left( s_B - 1 \right)
  + 2.03 \left(s-1\right)^2 -1.82 \left(s-1\right)^3 .
\end{equation*}
This relation should {\it not} be used for precision work,
such as attempting to combine data fit with different packages.

\subsection{Comparing the color parameters from SNLS and low-z data}
\label{subsec:colorcompare}
\nobreak
For the $U-B$ color relation, we can split the sample into low-z and
SNLS data as a test of SN evolution. This results in the values given
in table~\ref{tbl:colorparams}.  There is some tension between the
values.  However, this only includes statistical errors, and
systematics are expected to be of similar importance.  The details of
our systematics analysis will be presented with our cosmological
results elsewhere, but it is useful to summarize some of the more
important contributions.  The most important uncertainties arise from
the uncertainty in the magnitudes of our flux standard (Vega) on the
Landolt system (which affect our $K$-corrections) and the uncertainty
in the calibration of the low-z data.  The former are dominated by the
raw observational uncertainties of Vega from
\citet{1953ApJ...117..313J}.  To determine how well calibrated the
low-z data is to the Landolt system, we compare photometry of the same
SNe obtained by different observers \citep[e.g., ][]
{1999AJ....117..707R, 2006AJ....131..527J}, and conclude that there
is a global zeropoint uncertainty of about 0.015 mag in $BV$ for each
survey.  For $U$ the situation is more complicated.  As noted
previously, observer frame $U$ data is more difficult to calibrate
than $BV$, so we double this value to 0.03 mag in this filter.
However, as it turns out the $B$ zeropoint uncertainty dominates over
$U$, at least partially because it affects both $U-B$ and $B-V$.

In order to test the effects of these systematics, we adjust the
relevant parameter (e.g.\ the $U$ zeropoint), then re-derive the
SiFTO model and color relations.  We then consider the change in
$a_{\mathrm{SNLS}} - a_{\mathrm{low-z}}$ to be the resulting systematic
in $a$, etc.  For the uncertainties in the magnitude of Vega, we
find $\sigma_a = 0.014, \sigma_b = 0.050,$ and $\sigma_c = 0.007$.
We do not include $\sigma_{\mathrm{clr}}$ because its value is likely
to simply test the homogeneity of the data sample rather than be
a good test for evolutionary effects.  For the zeropoints of the low-z
data, we find $\sigma_a = 0.022, \sigma_b = 0.022,$ and $\sigma_c = 0.013$.
Other, smaller, contributions include the uncertainty in the MegaCam
and Landolt bandpasses and the MegaCam zeropoints.  The latter is
particularly important for $b$, giving $\sigma_b = 0.034$ for sub-1\%
zeropoint uncertainties.

Putting this information together, we conclude that
$\delta a = 0.179 \pm 0.112 \mathrm{(stat)} \pm 0.030 \mathrm{(sys)}$,
$\delta b = 0.157 \pm 0.152 \mathrm{(stat)} \pm 0.068 \mathrm{(sys)}$, and
$\delta c = 0.030 \pm 0.018 \mathrm{(stat)} \pm 0.019 \mathrm{(sys)}$.
The resulting \chisq\ is 4.57 for 3 degrees of freedom, which is
expected to occur by random chance in the case of no evolution 21\%
of the time.  

\section{HOW CLOSE TO PEAK DO WE NEED DATA?}
\nobreak 
\label{sec:daygap}
One question we can ask of our model is whether or not it can predict
the peak luminosity of the SN when we have no observations at that
epoch.  That is, how well can our model extrapolate from late time
observations to the peak?  In principle, the template error snake
should allow us to handle situations where there is little or no data
near peak, since the result will simply be a measurement with very
large errors.  However, the error snake is not known perfectly, so we
would like to test this statement. Most SN cosmology analyses have
required that SNe have data near peak, but the exact value used for
this cut varies from paper to paper and is usually not explained.  The
previous generation of SN fitting packages, which often did not
incorporate template errors, frequently produce implausibly precise
results when used in this situation.

Our test is based on real observations; a test on simulated data is
currently likely only to be useful as an internal consistency check.
We collect a sample of very well sampled, high-S/N light-curves with
data well before peak, and then, for each SN, compare the results of
the fits to all of the available data to those where we steadily
remove more and more photometry from around the peak.  We parameterize this
by the daygap, defined as the epoch of the first included point
relative to the date of $B$ maximum, and explore the range from $-$5 to
+15 days.  Our sample consists of the 11 low-z SNe: 1992A, 1995D,
1999aa, 1999dq, 1999ee, 2000E, 2002bo, 2002er, 2003du, 2004eo, and
2005cf.  Even these SNe generally do not have data before day $-$10,
which is why we have not tried to answer the complementary question of
how well our model can predict the peak magnitudes of SNe~Ia if we
only have pre-maximum observations.

There are two things we want to test: that the results are unbiased,
and that the errors reflect the missing data appropriately.  For the
latter, we compare the \chisq\ of the difference between the
light-curve parameters for the full light-curve, and for those at
various values of the daygap.  Since the different fits to the same SN
should be positively correlated, we will probably slightly
over-estimate our errors. This test is sensitive to small calibration
difficulties that arise when different telescope/detector systems
observe different parts of the light-curve, which is frequently the
case with low-z data.  This will artificially inflate the
differences. Therefore, the precise value should not be taken too
seriously; what we are looking for are sharp changes in the 
\chisq .

\begin{figure*}
\plotone{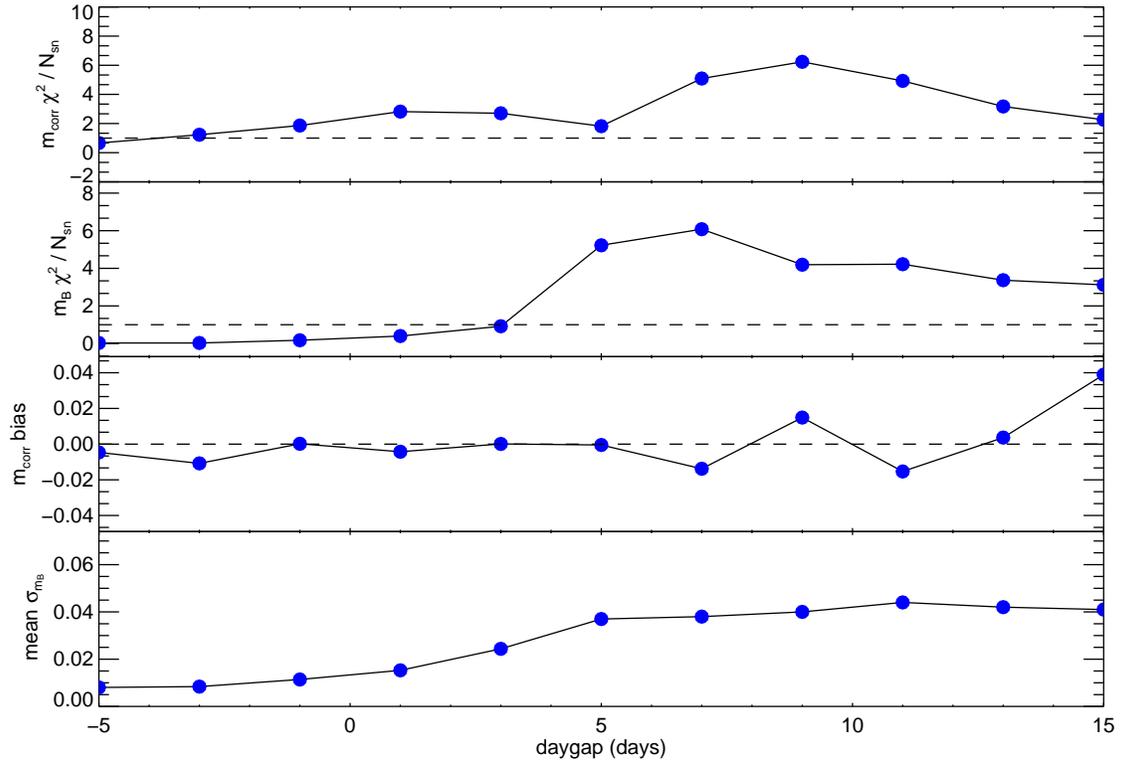}
\caption[Reduced \chisq\ for various parameters and values of
the daygap]{Reduced \chisq\ for the corrected magnitude
(equation~\ref{eqn:mbcorr}) as a function of the gap between
the first data point and the date of peak $B$ luminosity (top panel).
The middle panels show the same relation for the raw $B$ magnitude and
the bias in the corrected magnitude, and the bottom panel shows the
error in the $B$ magnitude. \label{fig:daygap}
}
\end{figure*}

The most important parameter to test is $m_{corr}$ as defined in
equation~\ref{eqn:mbcorr}, which is used to construct the Hubble
diagram.  Rough numerical simulation suggests that bias values larger
than about 0.02 are significant.  Most of the variables show
relatively little structure in their \chisq , so we only present the
two given in figure~\ref{fig:daygap}.  First, the \chisq\ of the peak
magnitude in $B$ rises sharply around day 5--10 after peak.  Second,
while this effect is also visible for $m_{corr}$, the effects are
slightly muted.  Our tests show little evidence for bias in the
corrected peak magnitude.  Based on the \chisq -like test, a good policy is
to require data within $\sim 7$ days of peak when using SiFTO.

Note that the accuracy of the derived parameters does degrade
considerably as the daygap is increased in all parameters (see the
bottom of figure~\ref{fig:daygap}), so SNe with no data near peak
will contribute relatively little information to the Hubble diagram.
We carry out the same test with other fitters in \S
\ref{subsec:lowzcomparison}, and find generally similar results.

\section{COMPARISONS WITH OTHER FITTERS}
\label{sec:comparison}
\nobreak 
In this section, we compare the results of SiFTO with those of other
fitters, particularly the published versions of SALT and SALT2. 
Our main focus will be on comparisons of SiFTO, SALT2, etc.\ when they
are used as distance estimators rather than as light-curve fitters.
There are a number of non-SN related settings that affect this comparison
(i.e., the assumed filter response of the Landolt
filters\footnote{SALT and SALT2 follow A06 in using blue-shifted
versions of the \citet{1990PASP..102.1181B} responses, while SiFTO
does not.}, etc.), so this comparison is not entirely fair.  In
\S~\ref{sec:retrain_compare} we present the results when these
differences are removed which provides a much more accurate test of
how different the results of the fitters really are.

We discuss MLCS2k2 in a more limited fashion in this section, keeping
in mind the following limitations: First, MLCS2k2 makes use of a
number of priors which should be accounted for in any detailed
comparison, and which in principle should be modified for the
particular data set under consideration.  Second, it fits in magnitude
space, which will introduce a bias when comparing low signal-to-noise
data\footnote{Subsequent to this writing, a version of MLCS2k2 which
works in flux space was kindly provided by G.~Narayan (2007, private
communication).  However, it does not address the other issues.}.
Finally, the current version of MLCS2k2 does not properly handle the
case where more than one observed filter maps to the same rest frame
Landolt filter.  This limits the sample size that can be compared
using SNLS data.

We also consider comparisons of the Hubble diagram.  All of the
fitters require that some additional uncertainty be added to the
luminosity distances when computing the best-fit cosmology.  This is
often referred to as ``intrinsic scatter'' or ``intrinsic
dispersion,'' and presumably reflects our imperfect understanding of
SN~Ia.  We denote this quantity by \sint , and find $\sint = 0.11$ mag
for SiFTO.  Comparing the values of \sint\ for the different fitters
is {\it not} currently a useful mechanism for comparison.  The issue
lies with the limitations of the template error snakes, which are
difficult to determine and are very sensitive to how accurate the
input flux errors are measured.  In addition, currently both SiFTO and
SALT2 do not include off-diagonal terms in the error snake (i.e.,
different epochs are considered uncorrelated), an assumption which is
clearly not true.  This is particularly an issue for SALT2, which
imposes the color relation during the fits.  Both effectively absorb
this unknown into \sint , and there is no reason to expect this to
affect both equally.  Therefore we limit ourselves to the more robust
comparison of the RMS of the resulting Hubble diagram.  However, the
RMS also includes the effect of measurement errors, so is only useful
as a comparison statistic if precisely the same set of input data are
used for all fitters.  A complication is that there is considerable
overlap between the data used to measure the RMS and that used to
derive the SiFTO model; the same issue applies to SALT2, MLCS2k2, and
(to a lesser extent) SALT.  However, we note that SiFTO/SALT/SALT2
make absolutely no use of relative distance information as part of the
training process.

\subsection{Comparison of SiFTO and SALT2 light-curve fits }
\label{subsec:lcfitcompare}
\begin{figure*}
\plotone{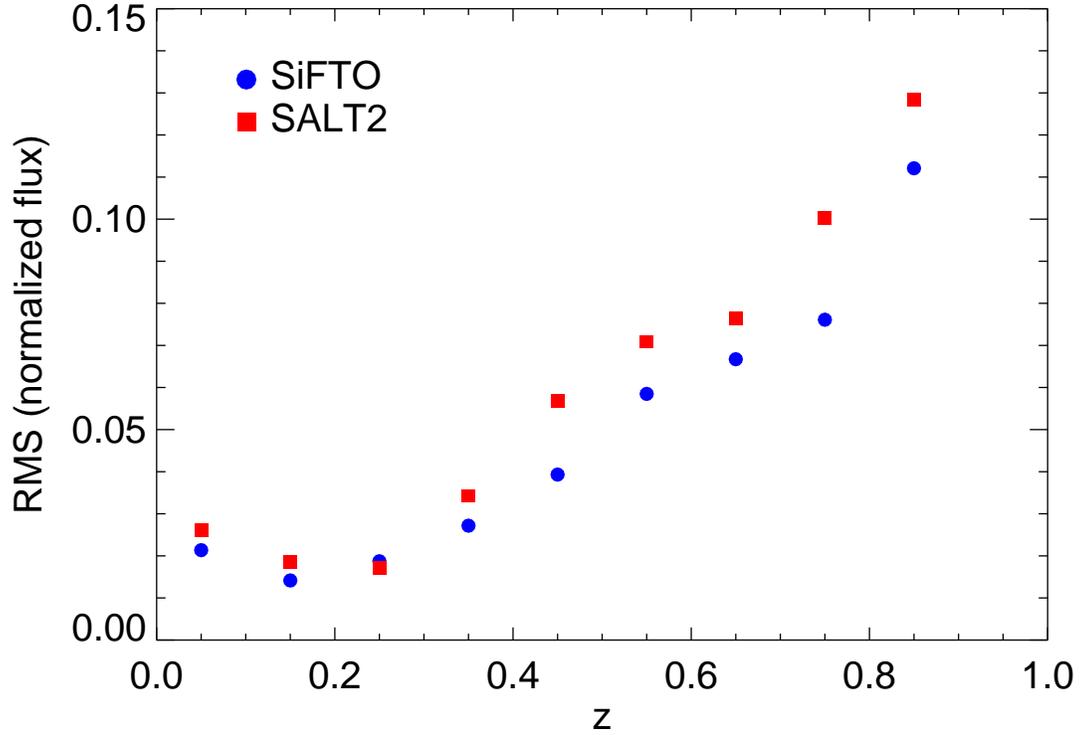}
\caption[Light-curve RMS comparisons]{Comparison of the RMS
 around the light-curve fits as a function of redshift for SALT2
 (red squares) and SiFTO (blue circles).  The residuals are normalized
 so that the model peak flux in each filter is unity. Note that the 
 SALT2 training sample includes the photometry used to perform this test, and
 that SiFTO has more free parameters in the fits.
 \label{fig:lcrms} }
\end{figure*}

\nobreak 
We begin by comparing SALT2 and SiFTO as light-curve fitters -- that
is, in terms of how well they fit the available photometry rather than
in terms of the precision of their relative distances.  Here we do
have the problem that the training process is specifically designed to
reduce the scatter of the input photometry with respect to the model,
so it is beneficial to train and test the models with disjoint data
sets.  We have carried this process out for SiFTO by randomly
splitting the training sample in half (giving about 90 SNe for both
testing and training), training a set of templates on the first half,
and then applying that model to the second half.  We then look at the
residual between the model and input flux normalized so that the
predicted peak flux in each filter is unity.  Because this includes
observational errors, we split this up into redshift bins ($\Delta z =
0.1$).  In order to mitigate photometry failures, we sigma-clip the
residuals at $4 \sigma$ for both fitters.  The results are shown for
SALT2 and SiFTO in figure~\ref{fig:lcrms} using the same photometry.
This comparison is unfair in two ways: first, the splitting into
disjoint training and testing samples has not been carried out for
SALT2.  Second, the SiFTO model has more free parameters since a color
relation is not imposed during the fitting process.  Nonetheless, the
two fitters give comparable results.

\subsection{Comparisons using low-z data}
\label{subsec:lowzcomparison}
\begin{figure*}
\plottwo{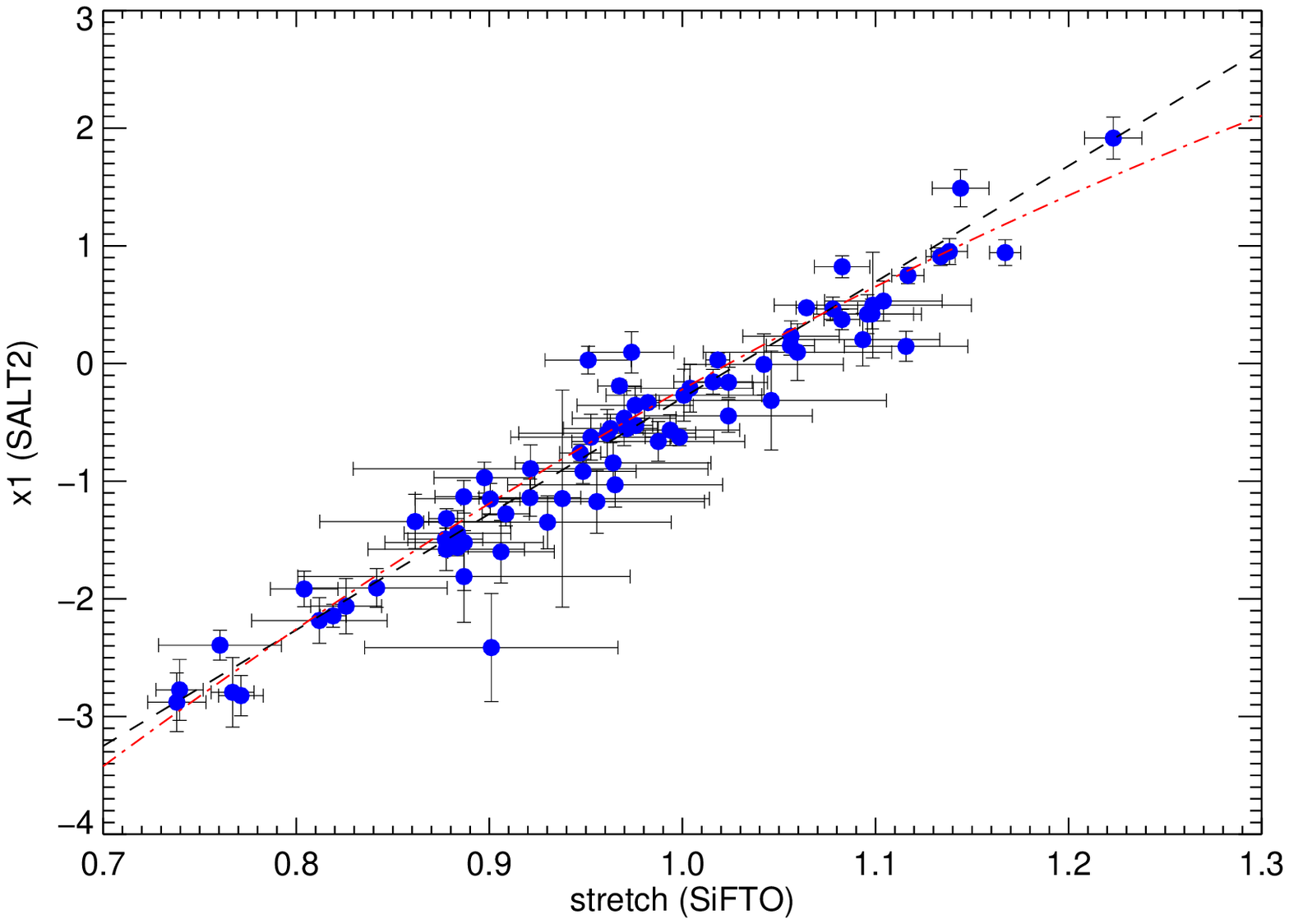}{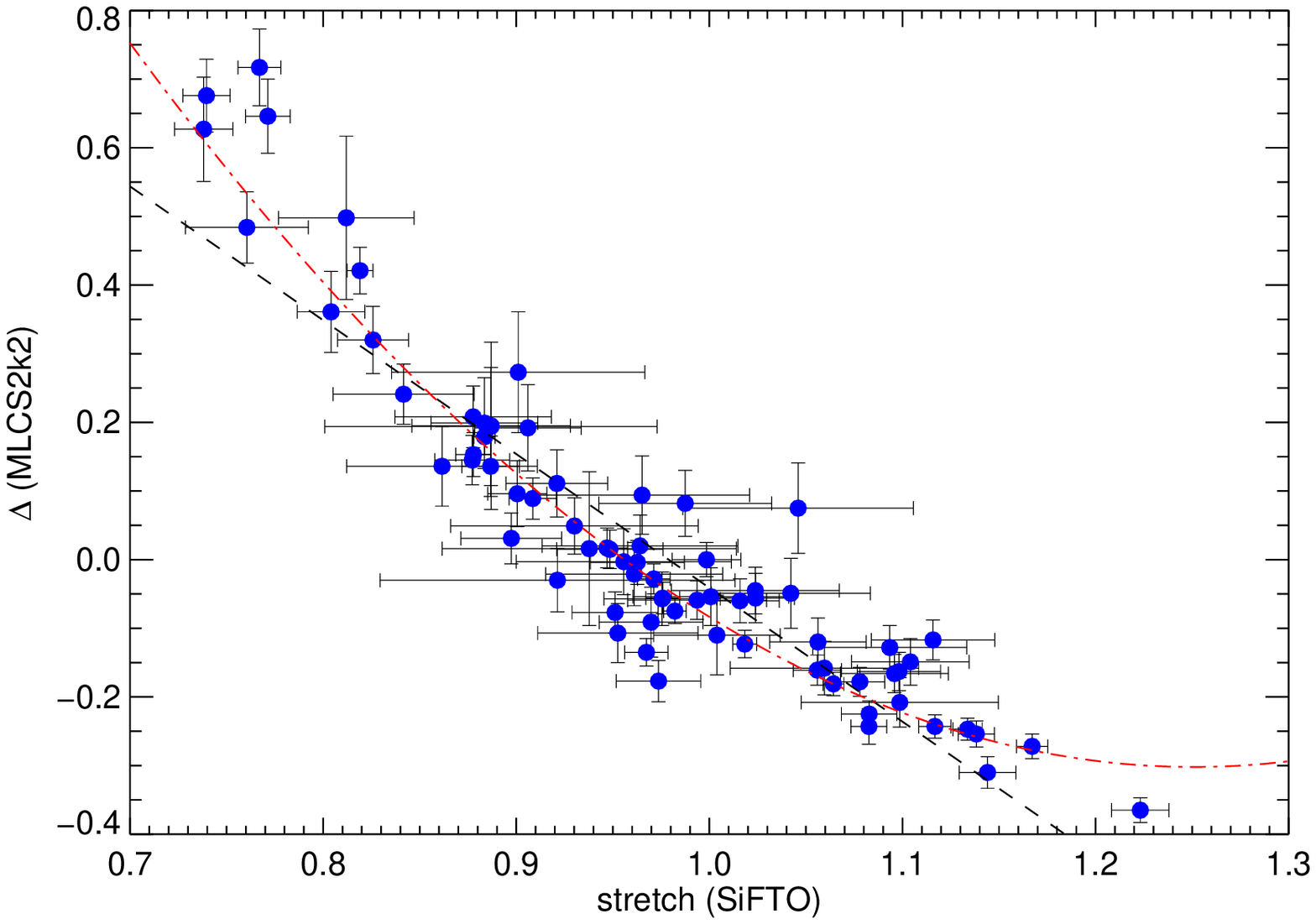}
\caption[Light-curve shape comparisons]{Comparisons of the light
 curve shape parameters for SiFTO vs.\ SALT2 (left) and MLCS2k2 (right).
 Note that the meanings of these parameters are quite different.
 Linear fits are given by black dashed lines, and 3$^{rd}$ order polynomial
 fits as red dot-dashed lines. \label{fig:stretch_compare} }
\end{figure*}

\begin{figure*}
\plottwo{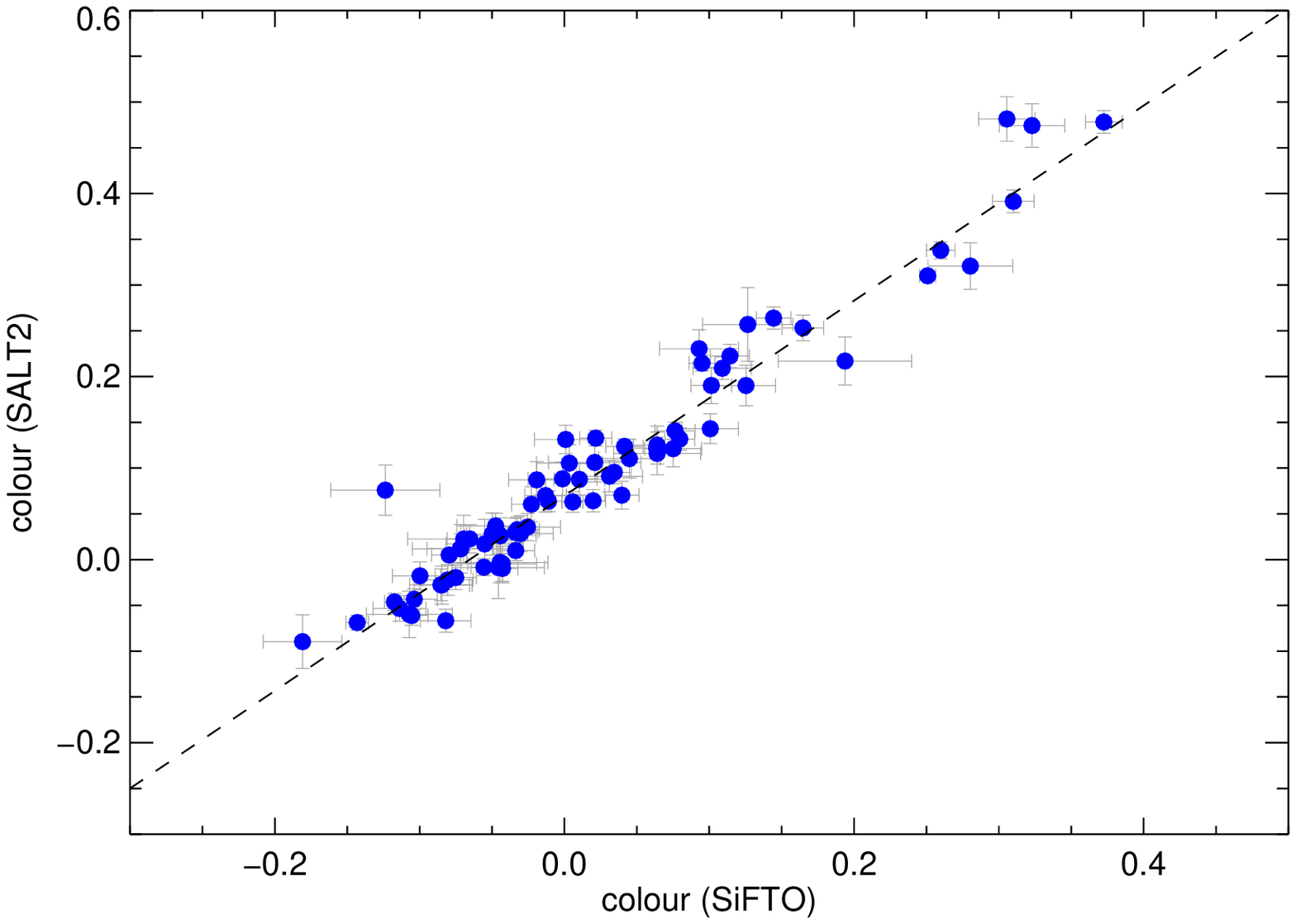}{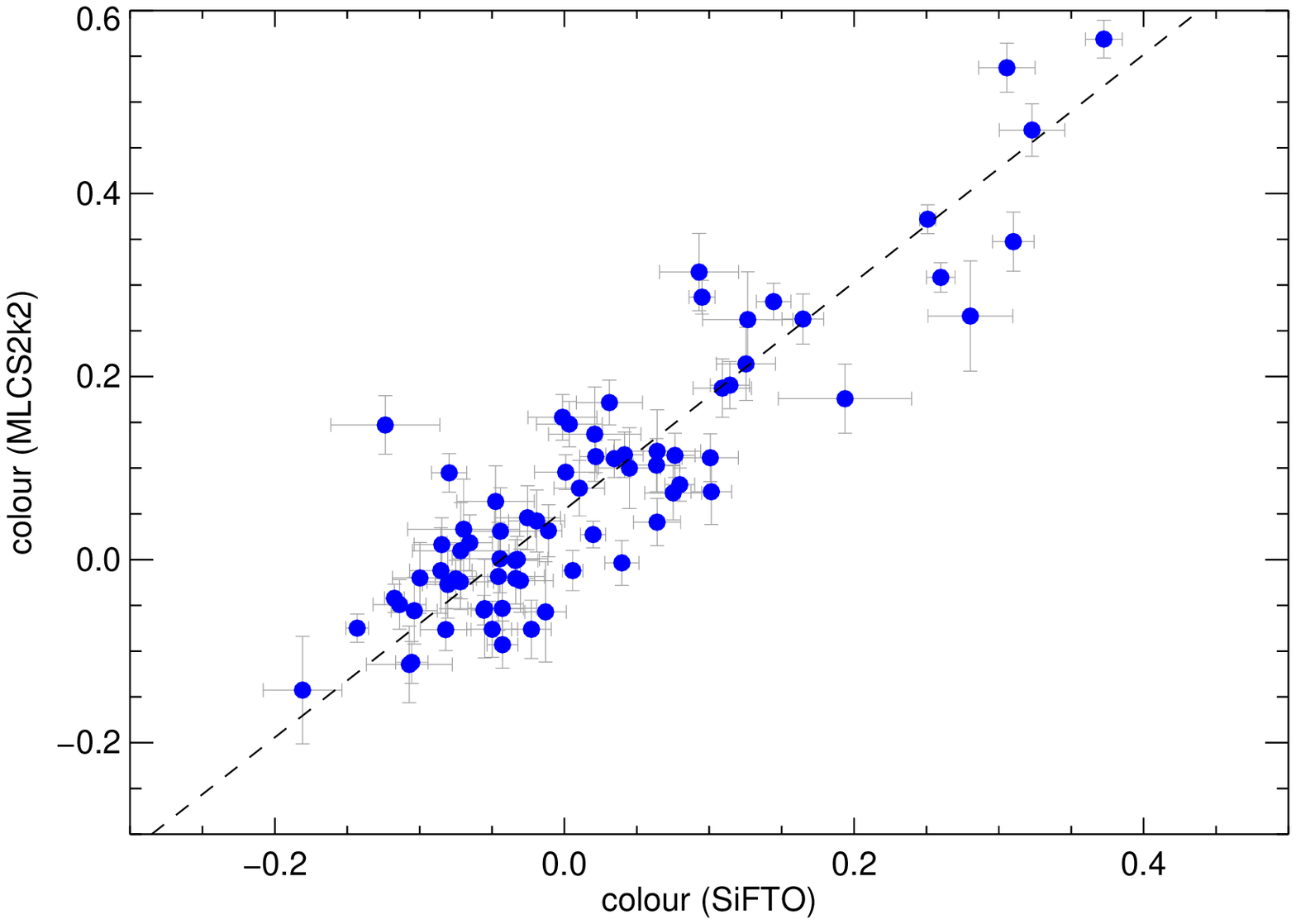}
\caption[Light-curve color comparisons]{Comparisons of estimated
 peak colors for SiFTO vs.\ SALT2 (left) and MLCS2k2 (right).
 The definitions of the peak color generally differ by a constant
 offset, so here we are testing if they are linearly related with
 a slope of one. The dashed lines are the best fit linear relations,
 which have slopes of $1.04 \pm 0.02$ and $1.2 \pm 0.03$, respectively.
 \label{fig:colour_compare} }
\end{figure*}

\nobreak 
The different light-curve fitters generally agree quite well on the
derived light-curve parameters, particularly when the MCLS2k2 priors
are removed.  Examples are shown in figures~\ref{fig:stretch_compare}
and \ref{fig:colour_compare}.  We generally do not expect perfect
agreement between the different light-curve parameters, but only that
there be a clear relation between the values.  This is particularly
true when discussing measures of the light-curve shape, which are
defined very differently, and SN color, which can have an arbitrary
offset related to the intrinsic color of the fiducial
SN~Ia.  The data set for this comparison is 78 low-z SNe from the
literature, using the same general requirements as were used in
\citet{2007ApJ...664L..13C}, although the sample here is larger
because we can include SNe which are not in the smooth Hubble flow.

The derived parameters are clearly quite correlated between the
different fitters, although there are a number of outliers in each
relation.  The $m_B$ comparison is consistent at about the 1\% level
globally, and since for most purposes only the relative distances matter,
this is encouraging.

Next we turn to the RMS around the low-redshift Hubble diagram.  We
take the above sample, remove all SNe with $z < 0.015$, and
then fit the Hubble diagram assuming $\om = 0.3$ and $\ol = 0.7$
-- the exact values are unimportant because of the limited redshift
range of our test sample.  The RMS values and \sint\ are given in
table~\ref{tbl:lowzvals}.  MLCS2k2 is not included in this test
because of complications related to the interpretation of SN colors,
which in turn is related to the question of whether or not the low-z
Hubble flow is really smooth or has a Hubble bubble, 
something that MLCS2k2 favors and the other fitters do not
\citep{2007ApJ...664L..13C} .  Since we do not include a bubble in our test,
this will tend to artificially inflate the RMS for MCLS2k2.  However,
we do perform the test using a larger data set below, where this issue
is mitigated by a larger, higher-redshift sample.

\begin{deluxetable}{lll}
\tablecaption{ Comparison of Low-z Hubble Diagram Fits }
\tablehead{ \colhead{Method} & \colhead{RMS (mag)} &
 \colhead{ \sint\ (mag)}
}
\startdata
SALT  & 0.156 & 0.134 \\
SALT2 & 0.158 & 0.129 \\
SiFTO & 0.155 & 0.119 \\ 
\enddata
\tablecomments{ The RMS and \sint\ values for fits to
 the same low-z data, and assuming $\om = 0.3$, $\ol = 0.7$.
 \sint\ is not useful for evaluating the different fitters.
 \label{tbl:lowzvals} }
\end{deluxetable}

We find that the three fitters perform equally well on this data set.
This is simultaneously encouraging and discouraging; it gives us
confidence in the different fitters, but at the same time suggests
that it is not easy to significantly improve the quality of the
distance estimates for SN~Ia beyond the $7-9\%$ level without some
improvement in the training data set.  Interestingly, when we perform
the same comparison with high-z SNe, SALT2 and SiFTO do considerably
better than SALT.

\begin{figure*}
\plotone{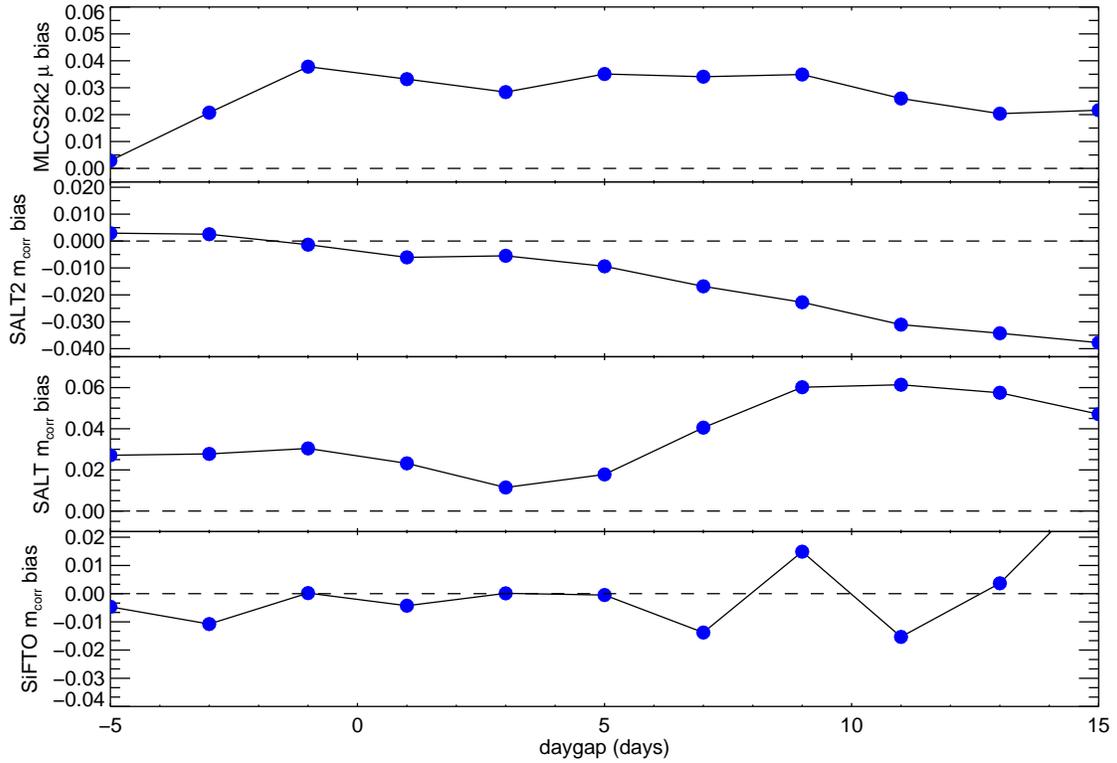}
\caption[Bias in the corrected peak magnitude as a function of the
 daygap]{Bias in the corrected peak magnitude relative to the case
 when all available data are used vs.\ the daygap for SALT, SALT2,
 and SiFTO, as well as for the equivalent MLCS2k2 variable ($\mu$).
 The error in this measurement is $\sim 0.015$ mag.
 \label{fig:daygap_bias} }
\end{figure*}

\begin{figure*}
\plotone{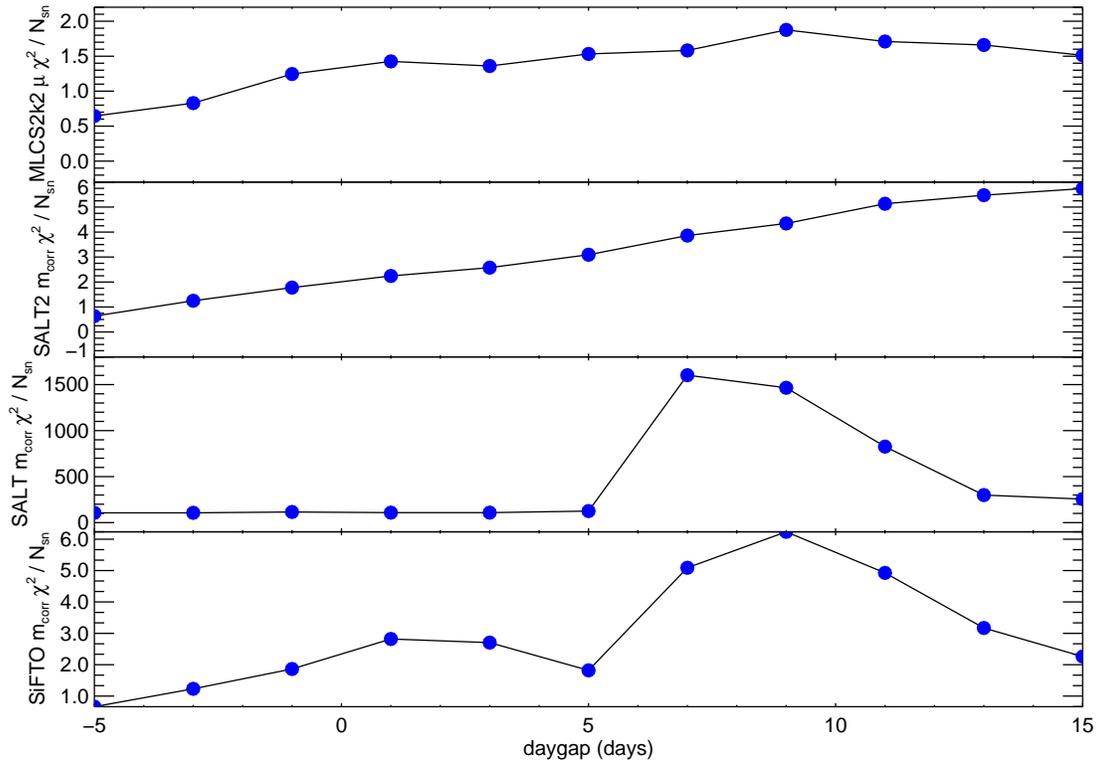}
\caption[Reduced \chisq\ for $m_{corr}$ versus the daygap
 for SALT, SALT2, SiFTO, and MLCS2k2.]{Reduced \chisq\ for the
 corrected peak magnitude $m_{corr}$ (i.e., the value used to
 construct the Hubble diagram) vs.\ the daygap for SALT, SALT2, and
 SiFTO.  Also plotted is the reduced \chisq\ for the MLCS2k2 distance
 modulus $\mu$, which is roughly
 equivalent. \label{fig:daygap_compare} }
\end{figure*}

Finally, we can test the other fitters to see how well they fare if
there are no data near peak, using the same method and set of 11 SNe as
in \S~\ref{sec:daygap}.  It is particularly interesting to consider
SALT, which alone among those discussed here does not have a template
error snake.  The bias values are shown in
figure~\ref{fig:daygap_bias} and the \chisq -like test results in
figure~\ref{fig:daygap_compare}.  MLCS2k2 performs the best in the
\chisq -like test, which is also true for the individual fit
quantities. MLCS2k2 and SALT display some bias even for low values of
the daygap.  One possible explanation is that both fitters would
benefit from some improvement in their pre-maximum light-curve model.

We note that SALT does quite poorly in the \chisq -like test, as
expected.  Furthermore, we obtain similar results when we do not use
the template error snake in SiFTO. The real lesson is then
that including template errors plays a critical role in
obtaining reliable fits when there are no data near peak.

\subsection{Comparisons using high-z data}
\nobreak
\label{subsec:hizcomparison}
It is also interesting to compare the results of fits to high-redshift
SNLS data.  We first compare SiFTO to the published versions of SALT
and SALT2.  Because we can apply each fitter to the same photometry,
again the best comparison statistic is the RMS of the resulting Hubble
diagram around the best fit.  The alternatives all have problems.  In
particular, comparing the resulting errors on the cosmological
parameters is a badly flawed metric unless the fits are virtually
identical, given the non-linearity of the parameter space.  An
additional complication occurs if the constraint contours impinge on a
physically prohibited region, such as $\om < 0$.  Tests with synthetic
data sets are perhaps useful as an internal consistency check, but we
do not currently understand SNe~Ia well enough for them to have any
absolute meaning.

For this set we start with the set used in A06 and modify it slightly.
We removed two SNe without solid spectroscopic type identification
(04D2iu and 04D3is), one with photometry problems (03D4cn), and one
which is probably SN 1991bg-like, and hence not well represented by any
of the models (SN 1992bf).  We also removed a number of SNe whose
light-curves are poorly sampled (SNe 1997Y, 1998eg, 03D3bh,
03D4gl, 04D1ag).  Finally, we added two low-z objects, SNe 2001V and
2002bf.  This leaves us with a sample of 43 nearby SNe and 65
high-redshift ones.

We carry out three cosmological fits for each: a flat-Universe \om\
fit, a \om\ + \ol\ fit, and a flat-Universe \om + $w$ fit.  For the
latter, we incorporate the SDSS baryon acoustic oscillations (BAO)
measurements of \citet{2005ApJ...633..560E}, since these are
marvelously orthogonal to the SN constraints.  The results are given
in table~\ref{tbl:rmsvalues}.  A bootstrap analysis indicates that the
error in the RMS for this sample is $\sim 0.005$ mag.  From the
perspective of the RMS test, SiFTO and SALT2 are indistinguishable,
and both considerably outperform SALT.  We also demonstrate the
effects of not incorporating the $U_{02}$ filter in SiFTO, which worsens the
RMS values at approximately the $2 \sigma$ level.  We note that the
benefits of extending the model bluer than $U$ are greater for the third
year SNLS sample than are shown here.  This is probably a result of
the fact that the \zmeg\ exposure times have been increased, so our $z
\ge 0.8$ SNe now have better measurements of $m_B$ and hence have
increased weight in our fits.  The RMS including high-z data is
slightly worse than if only low-z SNe are used, which simply reflects
the larger observational errors of the distant sample.

\begin{deluxetable}{lll}
\tablecaption{ Comparison of Fits to A06 Photometry }
\tablehead{ \colhead{ Fit type } & \colhead{ RMS } & 
  \colhead{ Parameters }
}
\startdata
\cutinhead{ SiFTO } \\
Flat \om       & 0.161 & $\om = 0.263 \pm 0.034$ \\
\om + \ol      & 0.160 & $\om + \ol = 1.358 \pm 0.353$ \\
               &       & $\om - \ol = -0.550 \pm 0.108$ \\
Flat \om + $w$ & 0.161 & $\om = 0.274 \pm 0.021$ \\
               &       & $ w = -1.047 \pm 0.088$  \\
\cutinhead{ SiFTO w/out $U_{02}$ }
Flat \om       & 0.171 & $\om = 0.261 \pm 0.035$ \\
\om + \ol      & 0.170 & $\om + \ol = 1.438 \pm 0.366$ \\
               &       & $\om - \ol = -0.568 \pm 0.110$ \\
Flat \om + $w$ & 0.172 & $\om = 0.273 \pm 0.022$ \\
               &       & $w = -1.054 \pm 0.088$  \\
\cutinhead{ SALT } \\
Flat \om       & 0.187 & $\om = 0.260 \pm 0.042$ \\
\om + \ol      & 0.187 & $\om + \ol = 1.023 \pm 0.388 $ \\
               &       & $\om - \ol = -0.477 \pm 0.120$ \\
Flat \om + $w$ & 0.188 & $\om = 0.272 \pm 0.022$ \\
               &       & $w = -1.030 \pm 0.098$  \\
\cutinhead{ SALT2 } \\
Flat \om       & 0.160 & $\om = 0.222 \pm 0.034$ \\
\om + \ol      & 0.160 & $\om + \ol = 1.306 \pm 0.350$ \\
               &       & $\om - \ol = -0.613 \pm 0.100$ \\
Flat \om + $w$ & 0.160 & $\om = 0.263 \pm 0.021$\\
               &       & $w = -1.126 \pm 0.090$  \\
\enddata
\tablecomments{ RMS values (in magnitudes) and resulting cosmological 
 parameters for fits to the A06 photometry using SiFTO, SALT, and SALT2.  Only
 statistical errors are included. The results of applying the SiFTO model
 limited to $\lambda > 3300 \AA$ (excluding $U_{02}$) are also shown.
 As explained in the text, only the RMS should be used as a performance 
 comparison statistic.  The RMS has an uncertainty of $\sim 0.005$ mag.
 \label{tbl:rmsvalues}
}
\end{deluxetable}

\begin{figure*}
\plotone{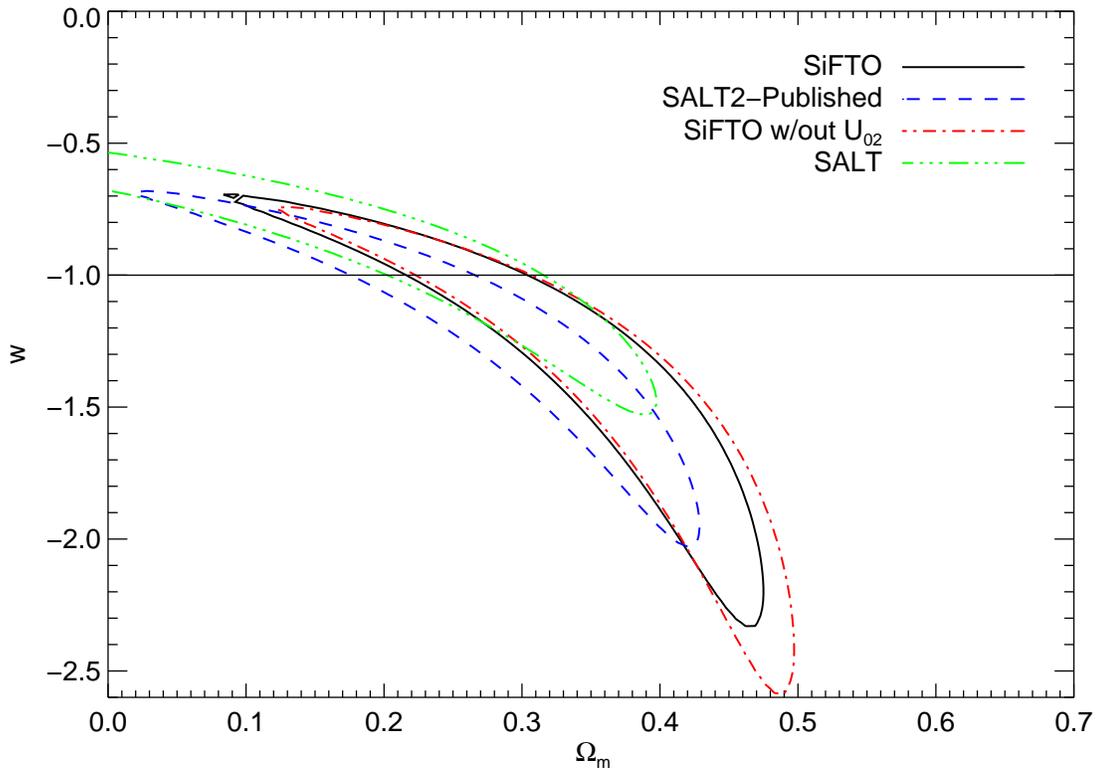}
\caption[Comparison of SiFTO, SALT, and SALT2 fits to the A06 photometry]
{Comparison of SiFTO, SALT and SALT2 fits to a subset of the A06 photometry.
 Only the 68.3\% confidence limits are shown for clarity. In addition,
 the constraints the result when $U_{02}$ is not included are also shown.
 \label{fig:fittercompare} 
}
\end{figure*}

Some of the resulting constraints are shown in
figure~\ref{fig:fittercompare}.  We see that some differences remain between
the fitters.  We shall turn to the question of whether this reflects
significant differences between the models in \S \ref{sec:retrain_compare}.
A weakness of the RMS as a test statistic is that it is sensitive to outliers.
Including $3 \sigma$ iterative outlier rejection does not
appreciably change the results.

We can perform a more limited version of this test including MLCS2k2,
although given the caveats given earlier this is somewhat less certain
and has a smaller sample size.  We obtain the results of
MLCS2k2 fits to the A06 photometry from \citet{2007ApJ...666..694W}.
This is a subset of the full data set, with about 93 SNe in total.
We therefore limit the SALT, SALT2, and SiFTO fits to the same sample.
We do not include the ESSENCE SNe in this test.  Carrying
out a similar analysis as the above, the RMS for SALT is 0.179 mag,
for SALT2 0.159 mag, for SiFTO 0.160 mag, and for MLCS2k2 0.205
mag. As noted for the low-z comparison, the MLCS2k2 value is somewhat
inflated by the issue of the Hubble bubble.  If we approximately remove
this effect by hand, the RMS for MLCS2k2 is about 0.19 mag, so we conclude
that its RMS is similar to SALT.  Adjusting the priors to better match
the survey properties of SNLS might improve the performance of
MLCS2k2.  It will be interesting to perform this comparison again once
the issues surrounding SN colors have been resolved.

\section{A BETTER COMPARISON BETWEEN SiFTO and SALT2}
\label{sec:retrain_compare}
\nobreak
The differences between the SALT2 and SiFTO results on the A06
photometry reflect not only differences in the underlying model
but also a variety of incidental settings, many of which are not
related to SNe at all.  While these differences are interesting, and play
a role in the systematics error budget of our final result, here they
simply obscure what we really want to know: what effect do the differences
in approach have on the cosmological parameters?

We can try to answer this question better by using the same non-SN
related settings for both fitters (filter functions, the magnitudes of
Vega on the Landolt system, etc.).  The published version of SALT2 was
trained on the set of SNLS photometry available at the time using the
MegaCam calibration of A06.  There are now more SNLS SNe available,
and the calibration has been improved.  We can further enhance the
comparison by training both fitters on exactly the same low-z and SNLS
photometry, using the same filter functions assumed by SALT2 in SiFTO, and
adding the additional low-z SNe provided by T.~Matheson to SALT2.

\begin{figure*}
\plotone{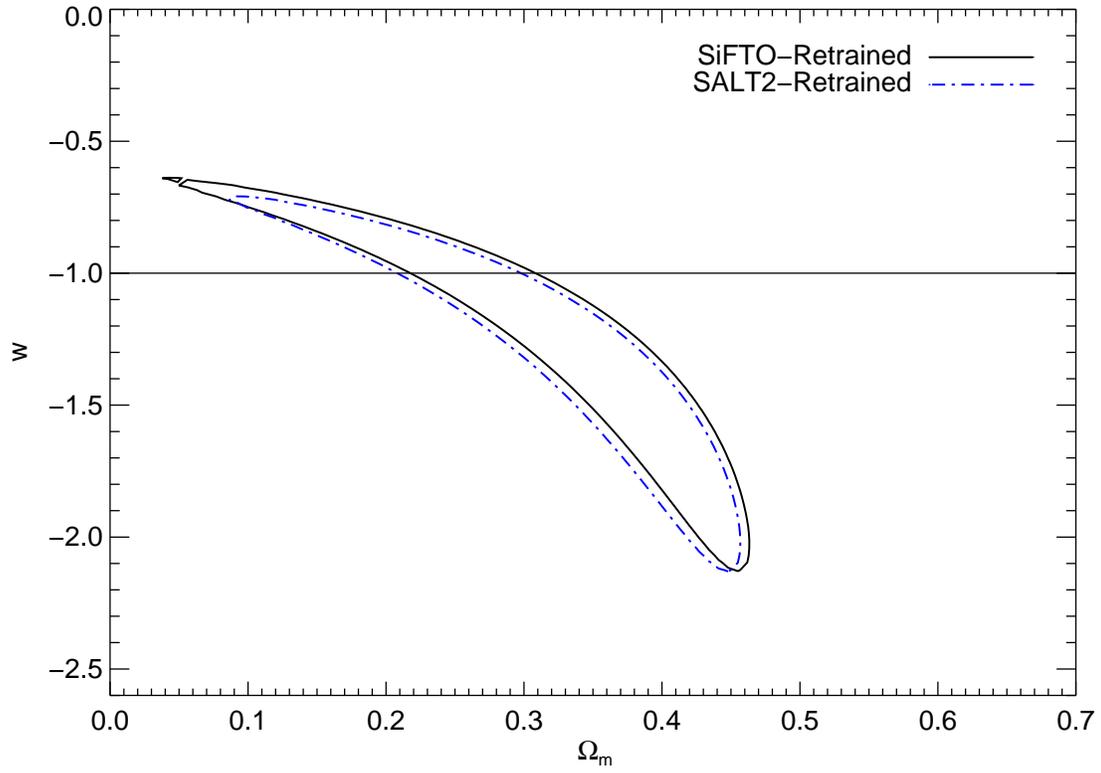}
\caption{Comparison of SALT2 and SiFTO with the same non-SN ancillary
settings and trained on the same photometry.  This is a better representation
of the fundamental differences between the models than that shown in
figure~\ref{fig:fittercompare}. \label{fig:fitterretraincompare} 
}
\end{figure*}

Carrying out this prescription, we obtain the results in
table~\ref{tbl:retrain_rmsvalues} and
figure~\ref{fig:fitterretraincompare}.  The results substantially
closer to each other.  This is very encouraging, and suggests that in
terms of the cosmological results the two approaches agree very well.

\begin{deluxetable}{lll}
\tablecaption{ Comparison of Fits to A06 Photometry with Retrained Models }
\tablehead{ \colhead{ Fit type } & \colhead{ RMS } & 
  \colhead{ Parameters }
}
\startdata
\cutinhead{ SiFTO } \\
Flat \om       & 0.159 & $\om = 0.262 \pm 0.033$ \\
\om + \ol      & 0.161 & $\om + \ol = 1.275 \pm 0.357$ \\
               &       & $\om - \ol = -0.536 \pm 0.106$ \\
Flat \om + $w$ & 0.160 & $\om = 0.272 \pm 0.021$ \\
               &       & $w = -1.042 \pm 0.086$  \\
\cutinhead{ SALT2 } \\
Flat \om       & 0.159 & $\om = 0.254 \pm 0.034$ \\
\om + \ol      & 0.160 & $\om + \ol = 1.393 \pm 0.353$ \\
               &       & $\om - \ol = -0.578 \pm 0.112$ \\
Flat \om + $w$ & 0.160 & $\om = 0.271 \pm 0.021$ \\
               &       & $w = -1.064 \pm 0.088$  \\
\enddata
\tablecomments{ RMS values (in magnitudes) and resulting cosmological 
 parameters for fits to the A06 photometry using versions of SiFTO and
 SALT2 trained on the same data with identical non-SN settings.  The
 SiFTO values are slightly different than in table~\ref{tbl:rmsvalues}
 because here SiFTO is using the MegaCam filter responses used by SALT2.
 \label{tbl:retrain_rmsvalues}
}
\end{deluxetable}

\section{CONCLUSIONS}
\label{sec:conclusions}
\nobreak 
We have described SiFTO, a light-curve analysis package developed for
use with SNLS data.  It is most similar in spirit to SALT/SALT2: all
three work by adjusting a model of the SED rather than with templates
in some rest-frame filters, and, like SALT2, SiFTO includes high-z
SNLS data in the training process.  This allows us to extend our model
blue-ward of the rest-frame $U$, which is extremely valuable when
fitting high-z SNe.  Distances and residuals from the Hubble relation
are not used in our training process, so this is not dependent on the
cosmological parameters.  The model can be applied to observations
from 2700 to 7100 \AA\ in the rest frame. 

The greatest difference between SALT/SALT2 and SiFTO is the manner in
which SN colors are handled.  SiFTO does not impose a color model
during the actual light-curve fit, but instead adjusts the SED to
match the observer frame colors.  The SED is then used to measure
various rest-frame colors, and these are combined to form a single
color parameter \scriptc .  The relation used to combine measurements
is derived using SNe at intermediate and low redshifts where each of
the colors are well measured.  The advantages of extending the model
further towards the near-UV, made possible by including hi-z data,
are significant.  This is an argument in favor of
the light-curve fitter class of packages (such as SALT, SALT2, and
SiFTO) versus distance estimators, which cannot incorporate
high-redshift data as easily.

We find that SiFTO produces the most reliable results when photometry
is available within 7 rest frame days of the $B$ peak, and data prior
to peak is not necessary.  The same result applies to the other
fitters as well.  Including a template error in the
model is particularly important when there are no data near peak.

We have carried out a number of tests comparing SiFTO to other
packages.  We find that SALT, SALT2, SiFTO, and MLCS2k2 agree fairly
well in terms of the derived light-curve parameters when the full SN
sample is considered.  In terms of their performance on the Hubble
diagram, for low-z data SALT, SALT2, and SiFTO are comparable.
However, when the test is extended to higher redshifts, SiFTO and
SALT2 out-perform SALT, and SiFTO performs as well as SALT2, while
MLCS2k2 performs about as well as SALT.  These comparisons show that
there are real, although small, differences in the derived
cosmological parameters from each when they applied to the same
photometry.  If we train SiFTO and SALT2 using the same non-SN related
settings and photometry, we find that the differences are much
smaller.  This indicates that, for currently available data, the
fundamental differences between the SALT2 and SiFTO models are minor
despite the significant differences in approach and algorithmic
design.

\acknowledgements 
The authors would like to thank Saurabh Jha and Adam Riess for 
making MLCS2k2 available, as well as for a number of useful discussions.
We would also like to thank Tom Matheson for providing additional low-z
spectroscopy prior to publication, and the anonymous referee for many
useful comments. M.~S.\ acknowledges support from the Royal Society.

\clearpage

\end{document}